\documentclass{emulateapj}

\begin{document}

\title{The Metallicities of Low Stellar Mass Galaxies and the Scatter in the Mass-Metallicity Relation}
\author{H. J. Zahid$^{1}$, F. Bresolin$^{1}$, L. J. Kewley$^{1}$, A. L. Coil$^{2}$, R. Dav\'{e}$^{3}$}
\affil{$^{1}$Institute for Astronomy, University of Hawaii, Manoa - 2680 Woodlawn Dr., Honolulu, HI 96822 \\
        $^{2}$Center for Astrophysics and Space Sciences, University of California, San Diego - 9500 Gilman Dr., La Jolla, CA 92093 \\
        $^{3}$Department of Astronomy, University of Arizona - 933 North Cherry Avenue, Tucson, AZ 85721}

\begin{abstract}

In this investigation we quantify the metallicities of low mass galaxies by constructing the most comprehensive census to date. We use galaxies from the SDSS and DEEP2 survey and estimate metallicities from their optical emission lines. We also use two smaller samples from the literature which have metallicities determined by the direct method using the temperature sensitive [OIII]$\lambda$4363 line. We examine the scatter in the local mass-metallicity (MZ) relation determined from $\sim$20,000 star-forming galaxies in the SDSS and show that it is larger at lower stellar masses, consistent with the theoretical scatter in the MZ relation determined from hydrodynamical simulations. We determine a lower limit for the scatter in metallicities of galaxies down to stellar masses of $\sim10^{7}$ M$_\odot$ that is only slightly smaller than the expected scatter inferred from the SDSS MZ relation and significantly larger than what is previously established in the literature. The average metallicity of star-forming galaxies increases with stellar mass. By examining the scatter in the SDSS MZ relation, we show that this is mostly due to the lowest metallicity galaxies. The population of low mass, metal-rich galaxies have properties which are consistent with previously identified galaxies that may be transitional objects between gas-rich dwarf irregulars and gas-poor dwarf spheroidals and ellipticals. 

\end{abstract}


\section{Introduction}

The metallicity of a galaxy is a key physical parameter for understanding its history and evolution. Starting with \citet{Pagel1979}, the practice of estimating gas-phase abundances from optical emission lines has allowed for significant advances in the study of galaxy evolution. Metals are produced in stars and returned to the interstellar medium (ISM) through stellar mass loss. In this way, the metallicity provides a record of the star formation history. However, the gas-phase abundance is modulated by gas flows where feedback processes can drive enriched gas out of galaxies and accretion of both primordial and enriched material can influence the gas-phase abundance. Observations of the relation between metallicity and other physical properties of galaxies and the evolution of these relations can place important constraints on the physical processes governing galaxy evolution.

The mass-metallicity (MZ) and luminosity-metallicity (LZ) relation was first observed by \citet{Lequeux1979} using nearby dwarf galaxies. Using $\sim$50,000 galaxies from the SDSS, \citet{Tremonti2004} have established the relation for the local population of star-forming galaxies down to stellar masses of log(M$_\ast) \sim$ 10$^{8.5}$ M$_\odot$. The MZ relation has also been observed at higher redshifts \citep[among others]{Erb2006b, Cowie2008, Lamareille2009, Mannucci2009, Zahid2011}. From the scatter in the MZ and LZ relation, many authors conclude that the fundamental correlation is between stellar mass and metallicity, such that lower mass galaxies have lower metallicity \citep[e.g.][]{Tremonti2004, Zahid2011}. \citet{Tremonti2004} attribute this effect to stellar wind driven mass loss enabled by the shallow potential wells in low mass galaxies. However, \citet{Dalcanton2004} argue that observations of the local MZ relation are also consistent with metal-poor infall of gas.

Several authors have suggested that gas flows alone cannot explain the MZ relation. Using smoothed particle hydrodynamical simulations, \citet{Brooks2007} have argued that in addition to mass loss driven by UV heating and supernova feedback, a mass dependent star formation efficiency can help to explain the origin and evolution of the MZ relation. In this scenario, due to lower star formation efficiencies, low mass galaxies have yet to convert much of their gas reservoir into stars and therefore are less metal-rich. Alternatively, \citet{Koppen2007} suggest that the MZ relation may result from variations of the initial mass function. They argue that in low mass galaxies the initial mass function does not extend to high stellar masses due to the low levels of star formation, thus resulting in fewer massive stars forming and less chemical enrichment.

The metallicities of low mass galaxies (M$_\ast$ $\lesssim$ 10$^{8.5}$ M$_\odot$) may provide important clues to the origin of the MZ relation which may help to distinguish between the various proposed scenarios. Many authors have investigated the metallicities of nearby dwarf galaxies \citep[among others]{Searle1972, Skillman1989, Kennicutt2001,  Lee2003, Hidalgo-Gamez2003, vanZee2006a, Zhao2010} but few have focused on the MZ relation \citep{Tamura2001, Lee2006, Vaduvescu2007}. \citet{Lee2006} extend the MZ relation by 2.5 orders of magnitude in stellar mass. Their main conclusion is that the scatter in the MZ relation is similar over 5 orders of magnitude.

Recent observations of low mass, metal-rich galaxies present a challenge to this conclusion \citep{Gu2006, Dellenbusch2007, Peeples2008, Petropoulou2011}. In particular, \citet{Peeples2008} focus on a population of metal-rich dwarf galaxies from the SDSS which they consider outliers to the MZ relation. They identify these galaxies as objects nearing the end of their star formation activity and transitioning from dwarf spirals and irregulars to dwarf spheroidals and ellipticals, consistent with the scenario discussed by \citet{Grebel2003}. However, citing possible systematic overestimates in the derived metallicities, \citet{Berg2011} refute this result. Due to the small number of observations, it remains unclear whether there is a population of low stellar mass, metal-rich galaxies. A population of galaxies, such as these, would present an important new test for chemical evolution models and cosmological simulations while providing stronger constraints for the physical mechanisms governing the MZ relation and galaxy evolution.

Here we examine the metallicities in galaxies spanning stellar masses from $10^{6} \lesssim$ M$_\ast$/M$_\odot$ $\lesssim 10^{10}$. In Section 2 we present the data and selection criteria used and in Section 3 we discuss our determination of stellar mass from photometry and metallicity from optical spectra. We examine the scatter in the MZ relation determined from $\sim20,000$ SDSS galaxies in Section 4 and in Section 5 we focus on the metallicities of low mass galaxies. In Section 6 we investigate systematic uncertainties in our mass and metallicity determinations. We present the main results of this study, the scatter in the MZ relation at low stellar masses in Section 7. The physical properties of these galaxies are examined in Section 8 and in Section 9 we provide a discussion. We give a summary of our results in Section 10. When necessary we adopt the standard cosmology $(H_{0}, \Omega_{m}, \Omega_{\Lambda})$ = (70 km s$^{-1}$ Mpc$^{-1}$, 0.3, 0.7).

\section{Data Sample}

\subsection{The SDSS Sample}

\begin{figure}
\includegraphics[width=\columnwidth]{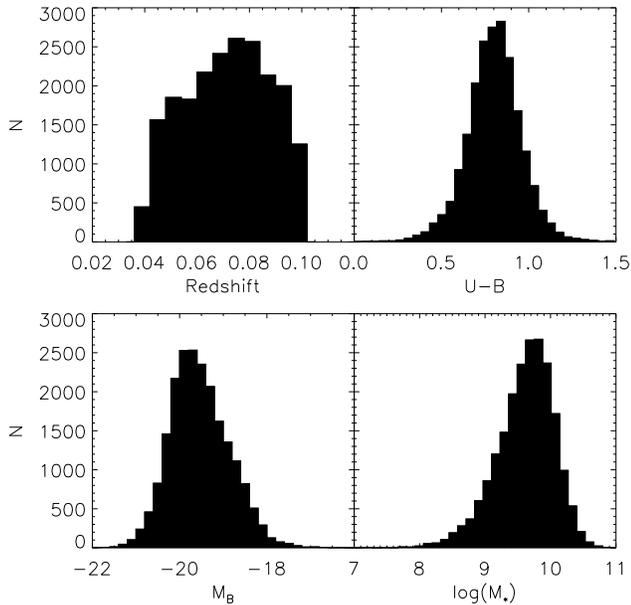}
\caption{The distribution of redshift (top left), U$-$B color (top right), absolute B-band magnitude (bottom left) and stellar mass (bottom right) for our selected sample of $\sim$20,000 SDSS galaxies. }
\label{fig:sel_sdss}
\end{figure}

The SDSS DR7 consists of $\sim$900,000 galaxies spanning a redshift range of $0 < z < 0.7$ \citep{Abazajian2009}. The survey has a Petrosian magnitude limit of $r_P$ = 17.8 and the spectra cover the nominal range of 3900 - 9100$\mathrm{\AA}$. For each object five bands of photometry, $ugriz$, is available \citep{Stoughton2002}. We use the publicly available emission line fluxes measured by the MPA-JHU group\footnote{http://www.mpa-garching.mpg.de/SDSS/DR7/}. We determine the local MZ relation from $\sim$20,000 galaxies using the sample selection of \citet{Zahid2011}. Hereafter, we refer to this relation as the SDSS MZ relation.

We first distinguish star-forming galaxies from AGN using the scheme of \citet{Kewley2006}. In order to avoid aperture effects, we require a g-band fiber aperture covering fraction $>30\%$ and impose a lower redshift limit of 0.04 \citep{Kewley2005}. The median covering fraction for the selected sample is $38\%$. \citet{Kewley2006} find that the SDSS sample is incomplete at higher redshifts and in order to minimize evolutionary effects we impose an upper limit redshift cutoff of z $=$ 0.1.

We require that the S/N H$\beta > 3$, $\sigma_{R23} < 2$ and equivalent width of H$\beta$ be greater than 4$\AA$. Here, $\sigma_{R23}$ is the error in the R23 parameter which is the sum of the flux in the [OIII]$\lambda4959,5007$ + [OII]$\lambda3727$ divided by the H$\beta$ flux. We find that the measured relation is not strongly dependent on the values adopted for these cuts and we select in this manner to be consistent with previous studies. In the selected sample, the median S/N of H$\alpha$ and H$\beta$ are 37 and 21, respectively.

Figure \ref{fig:sel_sdss} shows the distribution of redshift, U$-$B color, absolute B-band magnitude and stellar mass for our selected sample. This particular selection gives us a pure star-forming galaxy sample in a narrow redshift range. It covers 2 orders of magnitude in stellar mass and has a color distribution consistent with blue star-forming galaxies.

\subsection{Dwarf Galaxies Sample}

\begin{figure}
\includegraphics[width=\columnwidth]{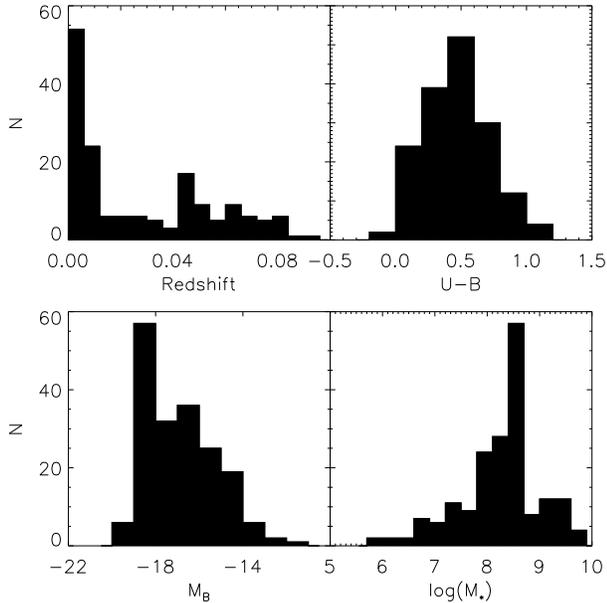}
\caption{The distribution of redshift (top left), U$-$B color (top right), absolute B-band magnitude (bottom left) and stellar mass (bottom right) for our sample of dwarf galaxies. For the 21 dwarf galaxies from \citet{Lee2006} we do not have redshift or color data.}
\label{fig:sel_dwarf}
\end{figure}

Many measurements of metallicities of dwarf galaxies can be found in the literature. We use the samples of \citet[see references therein]{Lee2006},  \citet{GilDePaz2003} and \citet{Peeples2008}. \citet{Lee2006} extend the SDSS MZ relation by 2.5 decades in stellar mass. Their sample is taken from the literature and consists of 27 dwarf galaxies. Using the models of \citet{Bell2001}, they derive masses from the $B-K$ color and scale them to a \citet{Chabrier2003} IMF. In this study we use the 21 galaxies in the \citet{Lee2006} sample for which there are metallicities determined by the direct method and we adopt their stellar mass determination.

\citet{GilDePaz2003} compile a sample of blue compact dwarf (BCD) galaxies. These galaxies are selected on basis of color, morphology and luminosity such that: $\mu_B - \mu_R < 1, \mu_B < 22$ mag arcsec$^{-2}$ and $M_K > -21$ mag. Here, $\mu_B$ and $\mu_R$ are the peak surface brightness in $B$ and $R$-band, respectively and $M_K$  is the K-band magnitude. \citet{Zhao2010} investigate the LZ relation for blue compact dwarfs using B and R-band photometry from this compilation with additional K$_s$-band photometry obtained from several sources \citep{Jarrett2000, Noeske2003, Noeske2005, Vaduvescu2007}. We determine, for the first time, the stellar masses from the photometry as described in Section 3.2. Of the 80 galaxies in the \citet{Zhao2010} sample, we select 66 with detections of [OIII]$\lambda4363$, allowing for metallicities to be determined using direct method. Similar to \citet{Lee2006}, the metallicities of the \citet{Zhao2010} sample are compiled from the literature and may suffer from aperture effects. However, by comparing abundances measured from several different sources in a subsample of the galaxies, \citet{Zhao2010} conclude that the effect is small. Both these samples with metallicities determined from the direct method reveal a population of low mass, metal-poor galaxies (see Section 5.1).

\citet{Peeples2008} examine the outliers from the mass-metallicity relation. Their sample consists of 41 metal-rich dwarf galaxies from the SDSS. We use the \citet{Peeples2008} sample but adopt the flux values of \citet{Berg2011} for the four galaxies they reexamined. Given the strict selection criteria of \citet{Peeples2008}, the ``very low mass" sample of galaxies (which have log(M$_\ast) \lesssim 9$) only consists of 17 galaxies. In order to investigate possible systematic effects in the determination of metallicities of low stellar mass galaxies, we supplement the \citet{Peeples2008} sample of 17 galaxies with 56 low mass, metal-rich galaxies taken from the SDSS sample described in section 2.1. We refer to these 56 galaxies as the supplemental SDSS sample. These galaxies have stellar masses $<$10$^{8.8}$M$_{\odot}$ and have high metallicities determined by three different diagnostics, making them outliers from the SDSS MZ relation. We use several independent methods of metallicity determination \citep[i.e.][see Section 3.1]{Yin2007, Kobulnicky2004, Kewley2002} in order to mitigate systematic effects that may result in overestimates of the metallicity with any one method.

Figure \ref{fig:sel_dwarf} shows the distribution of redshift, U$-$B color, absolute B-band magnitude and stellar mass for the dwarf galaxies sample. 

\subsection{The DEEP2 Sample}

\begin{figure}
\includegraphics[width=\columnwidth]{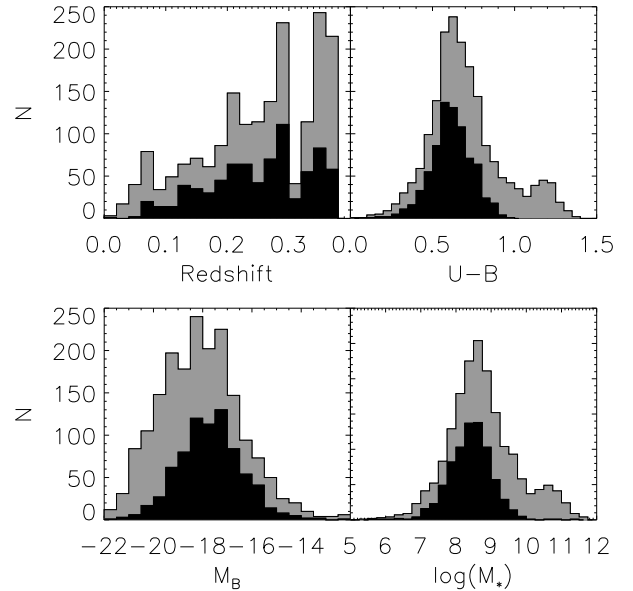}
\caption{The distribution of the parent (grey histogram) and selected (black histogram) sample of galaxies from the DEEP2 survey. The redshift distribution (top left) of the selected sample is roughly consistent with the parent sample. Our selection picks out the bluer (top right), less luminous (bottom left) and less massive (bottom right) galaxies as compared to the parent sample.}
\label{fig:sel}
\end{figure}

We probe significantly fainter objects than is possible with the SDSS by analyzing galaxies from the DEEP2 survey \citep{Davis2003}. This allows us to probe deeper to find low stellar mass objects otherwise not observed in shallow broadband surveys. The majority of galaxies in our parent sample come from observations of the Extended Groth Strip. The DEEP2 team used the DEIMOS multi-object spectrograph on the 10m Keck telescope to target galaxies in 4 fields covering 3.5 square-degrees down to a limiting magnitude of $R_{AB} = 24.1$. They primarily target galaxies in the redshift range of $0.7<z<1.4$. In order to maximize efficiency of finding high redshift galaxies, a color preselection approximately given by: B-R $<$ 2.35$\times$(R-I) - 0.45, R-I $>$ 1.15, or B-R $<$ 0.5 was applied for galaxies in fields 2-4. Galaxies in the Extended Groth Strip region (field 1) were targeted without a photometric redshift color pre-selection \citep{Coil2004}. 

The moderately high-resolution (R$\sim5000$) spectra have a nominal spectral coverage of $6500-9100\mathrm{\AA}$. BRI-band photometry for these galaxies was obtained by the DEEP2 team using the CFH12K camera on the 3.6 m Canada France Hawaii Telescope \citep{Coil2004} and for 54\% of our selected sample of galaxies $K_s$-band photometry is available from the Hale telescope at Mt. Palomar \citep{Bundy2006}.  In this study, we use the third release of the data\footnote{http://deep.berkeley.edu/DR3/}.

In order to estimate metallicities, we use the [NII]$\lambda6584$ and H$\alpha$ emission lines. These lines are observed in the DEEP2 data in the redshift range of $0<z<0.38$. We measure the uncalibrated flux of these emission lines by taking the area of a least-square gaussian fit. In many cases, the [NII] line can be very weak (S/N $<$ 3), so we simultaneously fit two gaussians with a single gaussian line-width to both emission lines. We use the same method to measure the [SII]$\lambda6717, 6731$ lines. We also measure the equivalent widths using the method outlined in the appendix of \citet{Zahid2011}. The error in the fitted line flux and equivalent width comes from propagating the measurement uncertainties in the spectrum. The median signal-to-noise (S/N) in our sample for H$\alpha$ is 22.

From the parent sample of 2,065 galaxies within the requisite redshift range, we select 777 galaxies for this analysis. In our selected sample, 695 are found in the Extended Groth Strip (field 1). By examining the photometric imaging, we find that 51 of the sources are star-forming knots in larger galaxies; these sources are removed from the sample. We select galaxies that have a reduced-$\chi^2$ of the fit less than 2 and a S/N of H$\alpha$ greater than 3. We also require that the $68\%$ confidence interval of our stellar mass estimate be smaller than 0.3 dex. These selections remove 1172 galaxies from the parent sample. Initially, we apply no S/N cut on the [NII]$\lambda6584$ line in order to avoid biasing against low metallicity objects. We remove 5 galaxies from the sample that have log([NII]/H$\alpha$) greater than maximum star-forming ratio of -0.3 \citep[e.g.][]{Kewley2006}. Following \citet{Weiner2007} who suggest that a substantial fraction of red emission line galaxies in the DEEP2 survey may be AGN contaminated, we remove 60 red galaxies using the blue and red galaxy color division for DEEP2 galaxies from \citet{Willmer2006}.

Figure \ref{fig:sel} shows the properties of the parent (grey histogram) and selected sample (black histogram). The redshift range of the parent and selected sample are comparable. Similar to our selection of SDSS galaxies, our selection of the DEEP2 data picks out the late-type blue star-forming galaxies which are generally less bright and less massive than early-type red galaxies. The U$-$B, M$_{B}$ and mass distributions in Figure \ref{fig:sel} reflects our selection.

\section{Methods}

\subsection{Mass Determination}

We estimate galaxy stellar masses by comparing photometry with stellar population synthesis models in order to determine the mass-to-light ratio which we use to scale the observed luminosity \citep{Bell2003a, Fontana2004}. Magnitudes are synthesized from the stellar templates of \citet{Bruzual2003} and an IMF described by \citet{Chabrier2003}. The 27 models span seven exponentially decreasing star formation models (SFR $\propto e^{-t/\tau}$) with $\tau = 0.1,0.3,1,2,3,5,10,15$ and $30$ Gyrs and two metallicities (0.4 and 1Z$_\odot$). The stellar population ages range from 0 to 13 Gyrs and we apply the extinction law of \citet{Calzetti2000} allowing E(B$-$V) to range from 0 to 0.6. The median statistical error for the derived stellar masses, determined from propagating the uncertainty in the photometry, is 0.15 dex. Though systematic effects can be significant \citep{Drory2004, Conroy2009a}, in all the samples except for the 21 galaxies from \citet{Lee2006}, we have consistently measured the stellar masses giving us a robust relative measure. The U and B-band absolute magnitudes for galaxies in the SDSS, DEEP2 and \citet{Zhao2010} samples are synthesized using this same procedure.

Emission line contributions are accounted for by taking the \citet{Kennicutt1998b} relation between the SFR and the UV luminosity which is synthesized as part of the mass determination procedure. In our sample, making no correction for the emission line contributions does not alter our mass determinations significantly. We adopt the median of the mass distribution and take the 68\% confidence interval as a measure of the error. In \citet{Zahid2011} we compare this method with the one used by the MPA/JHU group to determine stellar masses of the SDSS galaxies. We find that the results differ by a constant offset of $\sim$0.2 dex and that the dispersion between the two methods is 0.14 dex.

\subsection{Metallicity Determination}

The galaxies used in this study are not homogeneously observed. We therefore require several methods for determination of metallicities which we discuss here. Furthermore, different systematic uncertainties are associated with the various methods of metallicity determination, so when possible we use several methods for the same galaxy in order to assess and mitigate these uncertainties (see Section 6.2). 

Metallicities for the dwarf galaxies sample from \citet{Lee2006} and \citet{Zhao2010} are determined using the direct method (Section 3.2.1). For the SDSS and DEEP2 galaxies we determine metallicities using the empirically calibrated method of \citet[][Section 3.2.2]{Yin2007}. To address systematic bias and uncertainties we apply both empirically (Section 3.2.2) and theoretically (Section 3.2.3) calibrated methods in determining the scatter in the metallicities of local star-forming galaxies from SDSS. We also derive empirical corrections for metallicities determined from the [NII]$\lambda6584$ using the empirically calibrated N/O ratio (Section 3.2.4 and Section 6.2.2).

\subsubsection{Electron Temperature Method}

The ``direct" or electron temperature (T$_e$) method relies on measurements of the temperature sensitive auroral lines (e.g. [OIII]$\lambda4363$, [N II]$\lambda5755$). These lines provide tight constraints on the electron temperature and allow for straightforward determination of the gas-phase abundance. The metallicities are calculated using the iterative scheme presented by \citet{Izotov2006}. Due to temperature fluctuations and gradients, the O$^+$ and O$^{2+}$ ions reside in different zones of the HII region. T$_e$([OIII]) is determined for the O$^{2+}$ ion from the [OIII]$\lambda4363$ line and a linear conversion is applied to determine the temperature in the region of O$^{+}$, T$_e$([OII]). Contributions from higher ionization states are considered negligible and the gas-phase abundance is taken as the sum of the abundance in the two zones. 

This method suffers from several drawbacks making its use impractical in many cases. The [OIII]$\lambda4363$ line is on the order of 100 times weaker than the strong oxygen emission lines (e.g. [OII]$\lambda3727$, [OIII]$\lambda5007$) observed in optical spectra thus necessitating high S/N measurements. In most cases, this method can only be applied to nearby galaxies where high S/N spectra are available. Due to efficient cooling of HII regions by metals, the line can only be observed in low metallicity HII regions ($< 0.5$ Z$_\odot$). Finally, in the presence of temperature gradients and fluctuations within HII regions, several authors have argued that the T$_e$ method may lead to underestimates of the metallicity \citep[e.g.][]{Stasinska2002, Stasinska2005, Bresolin2006, Pena-Guerrero2011}, although recent work by \citet{ Bresolin2009} and has shown good agreement between direct abundances of extragalactic HII regions and stellar metallicities for blue supergiants \citep[see also][]{Kudritzki2011}.

\subsubsection{Empirically Calibrated Method}

Semi-empirically calibrated methods rely on calibrating strong-line ratios to metallicities determined from the combination of the T$_e$ method along with detailed photoionization modeling. The [NII]$\lambda6584$ to H$\alpha$ ratio is shown to be strongly correlated to gas-phase oxygen abundance. \citet{Pettini2004} have calibrated this ratio to nearby HII regions. They have parameterized the linear relation as
\begin{equation}
\mathrm{12+log(O/H)} = 8.90 + 0.57 \times N2
\label{eq:met_pettini}
\end{equation}
where $N2$ = log([NII]$\lambda6584$/H$\alpha$). This relation is valid in the range of $-2.5<N2<-0.3$ and has $1\sigma$ intrinsic scatter of 0.18 dex.

The linear relation between $N2$ and metallicity parameterized by \citet{Pettini2004} is poorly constrained at metallicities below 12 + log(O/H) $\sim$ 7.7. \citet{Yin2007} place stronger constraints at lower metallicities on the relation between metallicity and $N2$ by using a sample $\sim700$ galaxies from the SDSS along with values found in the literature using only metallicities determined from the T$_e$ method, thus providing  a purely empirical calibration. They parameterize the linear relation as
\begin{equation}
\mathrm{12+log(O/H)} = 9.263 + 0.836 \times N2.
\label{eq:met_yin}
\end{equation}
Their relation is valid in the range of $-2.5<N2<-0.5$ and has a $1\sigma$ intrinsic scatter of 0.16 dex. In this study, we are investigating low mass galaxies in the local universe and therefore find it desirable to have a metallicity diagnostic appropriately constrained at low metallicities. When determining metallicities from $N2$, we adopt the parameterization of \citet{Yin2007}. In the appendix we compare this parameterization with \citet{Pettini2004} at low stellar masses.

\subsubsection{Theoretically Calibrated Method}

A third class of metallicity diagnostics relies solely on calibration of strong-line ratios using photoionization models. These methods are not susceptible to observational limitations imposed by empirical calibrations relying on the faint [OIII]$\lambda4363$ line. Therefore, the model metallicities are well constrained and the parameterization is well defined over a broad range. 

The methods of \citet{Kobulnicky2004} and \citet{Kewley2002} calibrate strong-line ratios to metallicities determined using photoionization models. \citet{Kobulnicky2004} rely on the $R23$ diagnostic, which is the sum of the flux of [OII]$\lambda$3727, 3729 and [OIII]$\lambda4959, 5007$ normalized to the H$\beta$ flux. Though \citet{Kobulnicky2004} apply explicit corrections for the ionization parameter variations, there still may be some systematic effects that are uncorrected. Therefore, we also use the \citet{Kewley2002} method which relies on the $N2O2$ ratio which is the [NII]$\lambda$6584 to [OII]$\lambda$3727, 3729 flux ratio. This diagnostic is shown to be independent of the ionization parameter. Finally, for the SDSS sample of galaxies we also use the \citet{Tremonti2004} method which relies on fitting theoretical models of integrated galaxy spectra to the most prominent optical lines. For a review of these methods, we refer the reader to the appendix of \citet{Kewley2008}. We provide conversions between two commonly used $R23$ diagnostics and the direct method in the appendix of this paper.

\subsubsection{Nitrogen to Oxygen Ratio}

In order to investigate the effect of enhanced nitrogen enrichment with respect to oxygen, we measure the nitrogen abundance from emission line fluxes. N/O is empirically calibrated using star-forming galaxies in the SDSS with direct measurements of ionic abundances by \citet{Amorin2010}. The calibration is given by
\begin{equation}
\mathrm{log(N/O)} =  -0.86 + 1.94 \times N2S2 + 0.55\times N2S2^{2},
\label{eq:no}
\end{equation}
where 
\begin{equation}
N2S2 = \mathrm{log}\left(\frac{\mathrm{[NII]}\lambda6584}{\mathrm{[SII]}\lambda6717, 6731}\right). 
\end{equation}
[SII] is strongly correlated to [OII] and due to less uncertainty in the reddening correction, the relation between N/O and [NII]/[SII] shows less dispersion ($\sim0.1$ dex) than the relation between N/O and [NII]/[OII]. 

\subsection{Ionization Parameter}

The ionization parameter characterizes the level of ionization of the gas in an HII region. The ionization parameter is defined as 
\begin{equation}
q = \frac{F_{i}}{n},
\end{equation}
where $F_{i}$ is the flux of ionizing photons per unit area and $n$ is the number density of hydrogen atoms. The ionization parameter has units of velocity and can be thought of as the maximum velocity attainable by an ionization front driven by the local radiation field. The ionization parameter can best be constrained observationally by comparing two ionization states of single species. A commonly used diagnostic for measuring the ionization parameter is the $O32$ parameter defined as
\begin{equation}
O32 = \mathrm{log}\left(\frac{\mathrm{[OIII]}\lambda4959, 5007}{\mathrm{[OII]}\lambda3727, 3729}\right).
\label{eq:o32}
\end{equation}
However, \citet{Kewley2002} have shown that the [OIII]/[OII] ratio is not only dependent on the ionization parameter, but is also sensitive to metallicity.

Conversely, many of the strong line diagnostics are sensitive to the ionization parameter. One approach to disentangling the relationship between metallicity and the ionization parameter is to assign an initial guess of the metallicity in order to constrain the ionization parameter. Using this estimate of the ionization parameter, a second, more accurate estimate of the metallicity can be obtained. This process can be iterated until the values of both the ionization parameter and metallicity converge.

We measure the ionization parameter and metallicity using the \citet{Kobulnicky2004} diagnostic. This method uses the iterative scheme described above. The ionization parameter is determined from the $O32$ line ratio and the metallicity is determined from the $R23$ line ratio, defined as
\begin{equation}
R23 = \mathrm{log}\left(\frac{\mathrm{[OIII]}\lambda4959, 5007 + \mathrm{[OII]}\lambda3727, 3729}{\mathrm{H}\beta}\right).
\label{eq:o32}
\end{equation}
For details of how metallicity and ionization parameter are determined see the appendix of \citet{Kewley2008}.

\subsection{Star Formation Rates}

The star formation rates (SFRs) for the SDSS samples were made available in the DR7. These SFRs are measured using the technique of \citet{Brinchmann2004} who empirically correct for aperture effects. \citet{Salim2007} found that in galaxies with low level star formation, the SFRs were overestimated using this technique. \citet{Salim2007} correct this overestimate and this has been applied to the SFRs provided in the DR7.

For the DEEP2 and \citet{Zhao2010} samples we use the conversion of \citet{Kennicutt1998b} to determine SFRs from H$\alpha$ luminosities. This is given by
\begin{equation}
\mathrm{SFR\, \,(M_\odot \,\, yr^{-1}) = 4.6 \times 10^{-42} \,\, L_{H\alpha} \,\, (ergs\,\, s^{-1})}, 
\end{equation}
where $\mathrm{L_{H\alpha}}$ is the H$\alpha$ luminosity and the pre-factor has been scaled down by a factor 1.7 to convert from a \citet{Salpeter1955} to a \citet{Chabrier2003} IMF. For the DEEP2 data we scale the uncalibrated H$\alpha$ fluxes to the photometry and for the \citet{Zhao2010} sample, H$\alpha$ luminosities are determined by \citet{GilDePaz2003} from narrow-band H$\alpha$ filter images. We correct for dust extinction in both these samples using the E(B$-$V) derived from the SED fit when determining stellar mass. 

\section{The SDSS MZ Relation and its Scatter}

\begin{figure}
\includegraphics[width=\columnwidth]{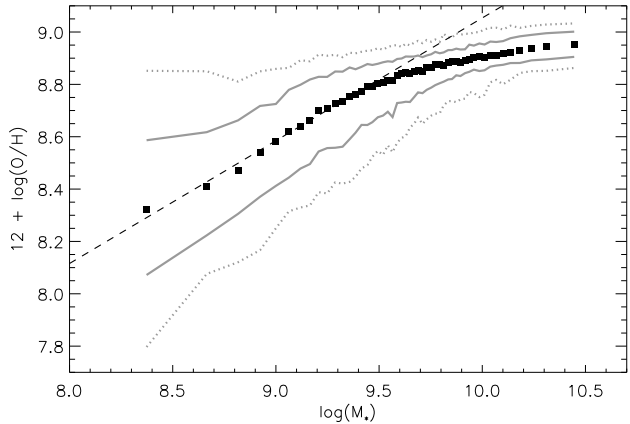}
\caption{The mass metallicity relation for SDSS galaxies. The black squares are median metallicity in 50 bins of stellar mass determined from $\sim$20,000 galaxies in the SDSS. The dashed line is a fit to the linear portion of the MZ relation (log(M$_{\ast}) \lesssim$ 9.4). The solid and dotted gray curves are the 68 and 95 percent contours of the distribution.}
\label{fig:mzr}
\end{figure}

We determine the SDSS MZ relation from $\sim$20,000 galaxies with metallicities determined using the $N2$ method given by Equation \ref{eq:met_yin}. In Figure \ref{fig:mzr}, we show the MZ relation and its scatter. We note that throughout this paper, when referring to the scatter in metallicities, we explicitly mean to refer to the range of observed metallicities at a fixed stellar mass. The relation is determined by sorting the data into 50 equally populated bins of stellar mass and taking the median mass and metallicity in each bin. The relation is plotted by the black squares and the scatter is shown by the solid and dotted grey curves which are the 68 and 95\% contours of the distribution, respectively. For galaxies at a fixed stellar masses $<10^{9.5} \mathrm{M}_\odot$ the scatter is symmetric. At higher stellar masses, the distribution is skewed towards lower metallicity galaxies (with a skewness near $-1$). A large part of this skewness can be attributed to the saturation of strong-line methods at high metallicities \citep{Kewley2002}. Additional skewness may also be the result of a physical upper limit to the metallicity attainable by a galaxy, though this is not well established and separating out these two effects is beyond the scope of this work.

The dashed line in Figure \ref{fig:mzr} is a linear fit to the data. The turnover observed in Figure \ref{fig:mzr} is due to a combination of a real turnover in metallicities as a function of stellar mass and saturation of the $N2$ parameter at high metallicities. We only fit to the linear portion of the SDSS MZ relation below the turnover (12 + log(O/H) $\lesssim$ 8.8). The fit is performed by minimizing the square of the residuals and is parameterized as
\begin{equation}
\mathrm{12 + log(O/H)} = (8.585 \pm 0.003) + (0.47 \pm 0.01) X_M.
\label{eq:mzfit}
\end{equation}
Here, $X_M = log(M_{\ast}) - 9$ and $M_\ast$ is the stellar mass in solar mass units. We fix the zero-point at a stellar mass of $10^9 M_{\odot}$ in order to reduce the covariance between the slope and intercept. The errors in the fit are assessed by bootstrapping the sample. We note that we consider this fitted MZ relation as the fiducial relation for later comparisons with the data.

\citet{Lee2006} examine the MZ relation for 27 low mass galaxies and conclude that the scatter is similar over five decades of stellar mass. However, using $\sim20,000$ galaxies from SDSS (Figure \ref{fig:mzr}), it becomes clear that the scatter increases substantially for lower stellar mass galaxies down to stellar mass of $\sim$10$^{8.5}$ M$_{\odot}$. We test whether the larger scatter at lower stellar masses is real or an artifact of the method used in determining metallicities. We examine the scatter as a function of stellar mass using several diagnostic methods of metallicity determination. We quantify the scatter in the MZ relation as the magnitude of the interval containing 95\% of the data for a given stellar mass bin. 

\begin{figure}
\includegraphics[width=\columnwidth]{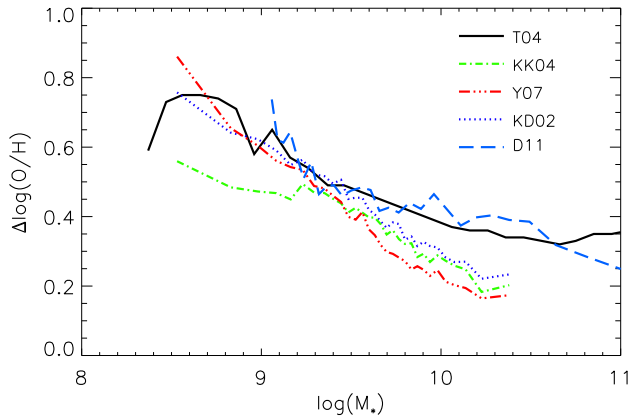}
\caption{The width of the metallicity interval containing 95\% of the data in each mass bin as a function of stellar mass for different metallicity diagnostics. The solid black line is the data taken from \citet{Tremonti2004}. The dot-dashed green curve, triple dot-dashed red curve and the dotted blue curve are the \citet{Zahid2011} selected data with metallicities determined using the \citet{Kobulnicky2004}, \citet{Yin2007} and \citet{Kewley2002} diagnostics, respectively. The blue long-dashed curve is the theoretical scatter from hydrodynamical simulations using a momentum conserving wind model \citep{Dave2011a}.}
\label{fig:scatter}
\end{figure}

Figure \ref{fig:scatter} shows the scatter as a function of stellar mass for several diagnostics. \citet{Tremonti2004} determine the MZ relation from $\sim$50,000 galaxies in the SDSS. The solid black line is the scatter in their determination of the MZ relation. We have subtracted 0.2 dex from their mass determination for consistency with our measurements of stellar mass \citep{Zahid2011}. The green dot-dashed, red triple dot-dashed and blue dotted curves are the scatter for the MZ relation, with the \citet{Zahid2011} selection, determined using the diagnostics of \citet{Kobulnicky2004}, \citet{Yin2007} and \citet{Kewley2002}, respectively. The blue long-dashed curve is the theoretical scatter determined from cosmological hydrodynamical simulations \citep[more details in Section 9]{Dave2011b}.

Strong line methods, in particular diagnostics using $N2$, are known to saturate at high metallicities. This is partly responsible for the low scatter observed at high stellar masses for the MZ relations determined from the $N2$ diagnostic (red curve). However, the trend of decreasing scatter in the MZ relation is continuous and is observed across all stellar masses, in particular at lower stellar masses where the strong-line diagnostics are not saturated. The mean relative errors in the $N2$, $N2O2$ and $R23$ line ratios used in the \citet{Yin2007}, \citet{Kewley2002} and \citet{Kobulnicky2004} diagnostics are 0.02, 0.05 and 0.04 dex, respectively. The observational uncertainties of the $N2O2$ and $R23$ line ratios increase with stellar mass due to the diminishing line strength of the oxygen lines at higher metallicity. Even at high stellar masses, the observational uncertainties are substantially smaller than the observed scatter and we note that for SDSS galaxies, systematic uncertainties associated with strong-line methods dominate over the observational uncertainties \citep{Kewley2008}. In general, the magnitude of the observed scatter in Figure \ref{fig:scatter} is significantly larger than either the observational or systematic uncertainties and we attribute this to intrinsic scatter in metallicities of galaxies at a fixed stellar mass.

In Figure \ref{fig:mzr} we see that the reason the scatter is larger at lower stellar mass is due to the fact that the metallicity of the most enriched galaxies in any given stellar mass bin decreases with a much more shallow slope than the least enriched galaxies. When comparing the high mass end to the low mass end of the MZ relation in Figure \ref{fig:mzr}, we see that the metallicities of the most enriched galaxies only decrease by $\sim$0.2 dex over 2 decades of stellar mass, whereas the metallicities of the least enriched galaxies decrease by $\sim$1 decade.

\section{Metallicities of Low Mass Galaxies}

In this section we extend our study of the scatter in the MZ relation to low mass galaxies. Because different methods have been applied in determining the metallicities of metal-poor and metal-rich galaxies, we present them separately in Sections 5.1 and 5.2, respectively.

\subsection{Metal-Poor Dwarf Galaxies}

\begin{figure}
\includegraphics[width = \columnwidth]{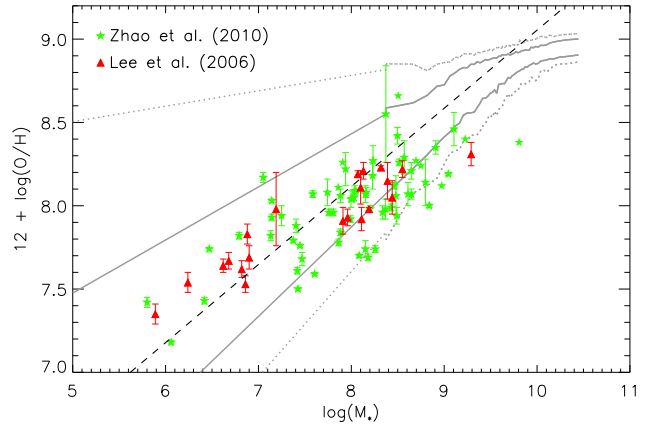}
\caption{The MZ relation for 21 dwarf irregular galaxies(red triangles) from the sample of \citet{Lee2006} and 66 compact blue galaxies (green stars) from the sample of \citet{Zhao2010}.}
\label{fig:sdss_lee}
\end{figure}

We plot the mass and metallicity of the sample of dwarf galaxies taken from \citet{Lee2006} and \citet{Zhao2010} in Figure \ref{fig:sdss_lee}. The metallicities for this sample have been determined using the direct method (see Section 3.2.1). The fiducial MZ relation is plotted as the dashed line. In Figure \ref{fig:sdss_lee} we linearly extrapolate the 68 (solid line) and 95\% (dotted line) contours to lower masses by fitting the contours in the linear portion of the fiducial relation (8.5 $\lesssim$ log(M$_\ast$) $\lesssim$9.4). We extrapolate the observed scatter to investigate whether at the low mass end the scatter is consistent with the expectation from the fiducial relation.

The slope of the MZ relation determined from this sample alone is $0.27\pm0.02$. The fit and error have been determined by a bootstrapping method. The slope differs significantly (9$\sigma$) from the slope of the fiducial relation ($0.47 \pm 0.01$, see Equation \ref{eq:mzfit}) determined from the $N2$ method as shown in Figure \ref{fig:mzr}. We note that the $N2$ method used in determining the fiducial relation is calibrated to be on the same absolute scale as the direct method. The metallicities of the 87 dwarf galaxies plotted in Figure \ref{fig:sdss_lee} deviate strongly from the distribution inferred from the SDSS MZ relation such that no low mass, metal-rich galaxies are observed. The smaller scatter observed in low stellar mass galaxies was noted by \citet{Lee2006}.

The small scatter observed in these data could be a result of selection bias attributable to the direct method of metallicity determination. The [OIII]$\lambda4363$ line becomes too weak to observe at $\sim$0.5 Z$_\odot$ (12 + log(O/H)$\sim$8.5) and above. This is because the [OIII]$\lambda4363$ line strength is anti-correlated to metallicity, such that lower metallicity objects have stronger emission. These observational effects could lead to an artificial suppression of the scatter. In particular, the fact that all the galaxies in the sample have 12 + log(O/H) $\lesssim$ 8.5 and that the highest metallicity galaxy observed at a given stellar mass is a strong function of stellar mass likely results from observational biases associated with the direct method.

\subsection{Metal-Rich Galaxies from SDSS and DEEP2}

\begin{figure*}
\includegraphics[]{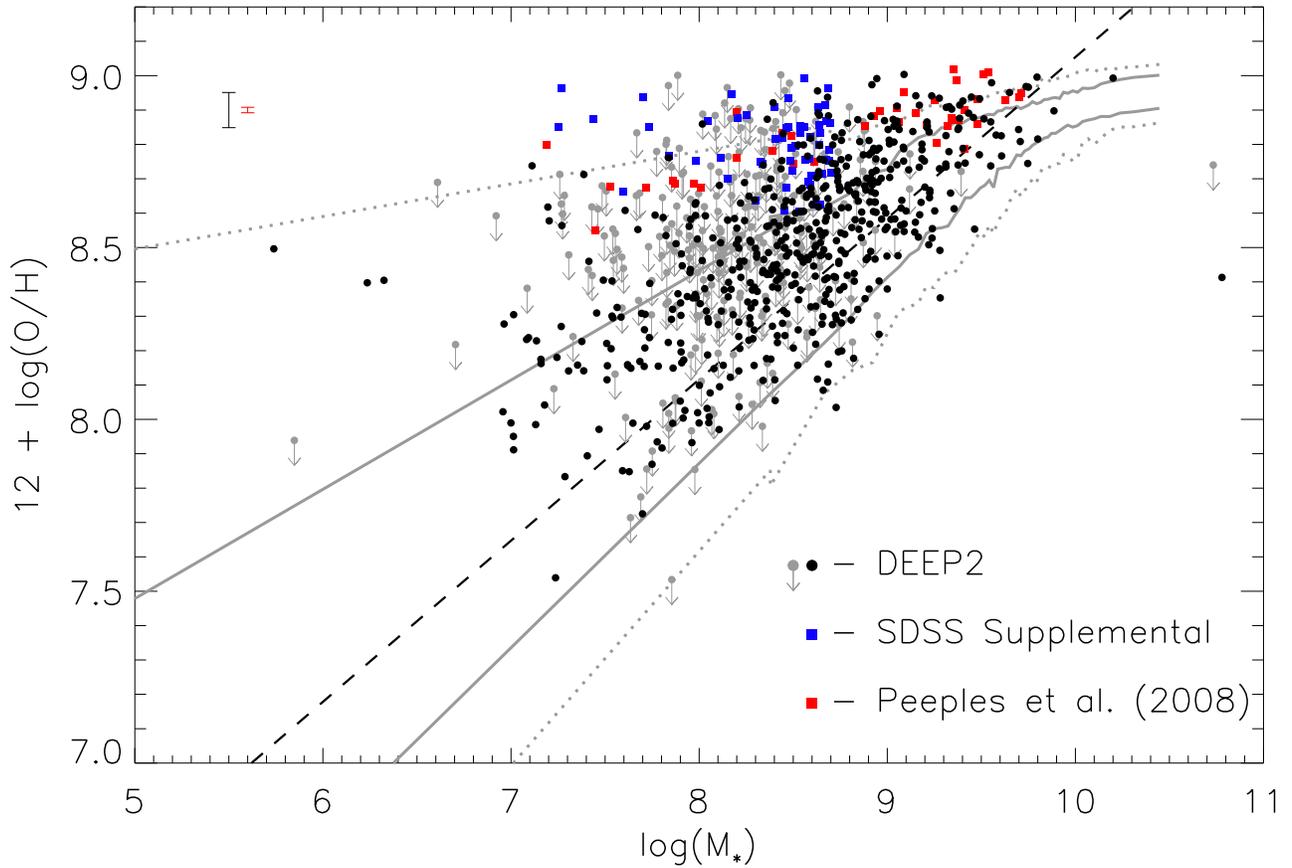}
\caption{The mass and metallicity for our sample of metal-rich galaxies. The black points are 534 galaxies from the DEEP2 survey. The gray arrows are an additional 248 galaxies from DEEP2 where we only measure upper limits. These upper limits are determined by adopting a 3$\sigma$ upper limit for non-detected [NII]$\lambda6584$. The red and blue squares are galaxies from the SDSS taken from \citet{Peeples2008} and our supplemental SDSS sample, respectively. The median error in the metallicities of the detected DEEP2 and SDSS sample of galaxies are 0.05 and 0.008 dex and are shown by the black and red error bars in the top left, respectively. The dashed line is the fit to the SDSS MZ relation take from Figure \ref{fig:sdss_lee}. The solid and dotted gray lines are the 68 and 95\% contours of the SDSS MZ relation. The fit and contours have been extended down to lower masses by linearly extrapolating the contour from the linear portion of the SDSS MZ relation (log(M$_{\ast}) \lesssim$ 9.4).}
\label{fig:deep}
\end{figure*}

In Figure \ref{fig:deep} we plot the masses and metallicities of the DEEP2 galaxies along with the sample from \citet{Peeples2008} and our supplemental SDSS sample. The metallicities for these galaxies are determined from the $N2$ calibration of \citet{Yin2007} given in Equation \ref{eq:met_yin}. For 31\% (240/770) of the DEEP2 sample of galaxies [NII]$\lambda6584$ is not detected at a 3$\sigma$ level. For these galaxies we adopt a 3$\sigma$ upper limit value for [NII]$\lambda6584$ when determining the metallicity. The dashed line is the extension of the fit to the linear portion of the SDSS MZ relation as seen in Figure \ref{fig:sdss_lee}. The 68 (solid line) and 95\% (dotted line) contours of the fiducial relation have been extrapolated for stellar masses below log(M$_\ast$) $\lesssim$ 8.5. The median observational uncertainty in the metallicities of the detected DEEP2 galaxies is 0.05 and for the \citet{Peeples2008} and SDSS supplemental sample it is 0.008 dex.

Opposite to the behavior of the [OIII]$\lambda4363$ line, the [NII]$\lambda6584$ line becomes stronger at higher metallicities. The fixed magnitude limit of the DEEP2 survey means that at lower stellar masses only the strongest [NII]$\lambda6584$ emitters will be be detected. This effect can be seen in Figure \ref{fig:deep} where the lowest metallicity object observed at a given mass increases in metallicity with decreasing stellar mass. This is opposite to the observational bias present in the samples of \citet{Lee2006} and \citet{Zhao2010} which have direct method metallicities.

These data for the first time reveal a population of low mass, metal-rich galaxies down to stellar masses of $\sim$10$^6$ M$_\odot$. By combining two methods in determining metallicity which have the opposite observational bias, we find a significantly higher scatter in the metallicities of low mass galaxies than previously reported.  Because of incompleteness and sample bias, we cannot reliably determine the MZ relation or quantify its scatter at low stellar masses. In Section 7 we derive a lower limit for the observed scatter in the metallicities of low stellar mass galaxies.

\section{Systematic Uncertainties in Mass and Metallicity}

\subsection{Stellar Mass Estimates}

Our apparently low mass, metal-rich objects could be higher mass galaxies with systematically underestimated stellar masses. This could possibly result from heavy obscuration due to dust. For our sample of 94 metal-rich galaxies from SDSS, we determine the extinction from the Balmer decrement. The mean E(B$-$V) for our sample is $0.29\pm0.12$. We do not observe H$\beta$ in our DEEP2 sample and therefore determine the extinction from the SED fit. The mean E(B$-$V) for our DEEP2 sample is $0.26\pm0.16$. Our metal-rich sample of galaxies have values of extinction consistent with star-forming galaxies in the local universe \citep[e.g.][]{Jansen2001} and are not found to suffer from heavy extinction. Moreover, our method for determining stellar masses corrects for extinction and we remove galaxies from our sample that have stellar mass estimates that have 68\% confidence intervals $>0.3$ dex.

\begin{figure}
\includegraphics[width=\columnwidth]{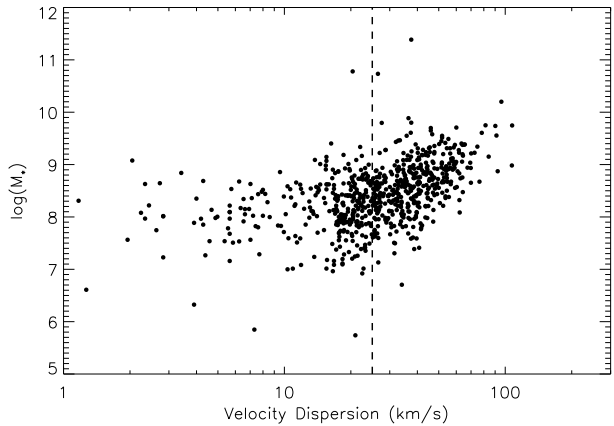}
\caption{The stellar mass plotted against the velocity dispersion for our sample of DEEP2 galaxies. The velocity dispersion below 25 km/s (dashed line) has large errors due to instrumental resolution.}
\label{fig:vdisp}
\end{figure}

The velocity dispersion of a galaxy has been shown to correlate with the stellar mass with some scatter \citep[e.g.][]{Kassin2007}. Therefore, low velocity dispersion can be taken as an indication of low stellar mass. The line-of-sight velocity dispersion for the DEEP2 sample is given by \citet{Weiner2006} as
\begin{equation}
\sigma_{disp} = \frac{c}{\lambda_{obs}} (\sigma_{obs} - \sigma_{inst})^{1/2}, 
\label{eq:vdisp}
\end{equation}
where $\sigma_{disp}$ is the velocity dispersion given in km s$^{-1}$, $\lambda_{obs}$ is the observed wavelength and $\sigma_{obs}$ and $\sigma_{inst}$ are the observed line width and instrumental resolution, respectively. For the DEEP2 sample, the velocity dispersion is primarily determined from the H$\alpha$ emission line. For the DEEP2 data the instrumental resolution is 0.56$\mathrm{\AA}$ \citep{Weiner2006}. The velocity dispersion becomes unreliable when $\sigma_{disp} > c \sigma_{inst}/\lambda_{obs}$ due to the fact that small errors in the line width translate to large errors in the velocity dispersion \citep{Weiner2006}. 

Figure \ref{fig:vdisp} shows that the stellar mass is correlated to the velocity dispersion, as expected. The median velocity dispersion of our DEEP2 sample is 26 km s$^{-1}$. Below $\sim$25 km s$^{-1}$, shown by the vertical dashed line, galaxies have unreliable determination of the velocity dispersion due to their small line widths but are still consistent with low stellar masses. We are not able to perform the same comparison for our sample of galaxies from SDSS because of the lower instrumental resolution of that survey.

\subsection{Metallicity Estimates} 

Recently, \citet{Berg2011} have reexamined four of the metal-rich galaxies from the sample of \citet{Peeples2008}. Using new spectroscopic observations along with detailed comparisons with photoionization models, \citet{Berg2011} conclude that the metallicities of the four galaxies have been overestimated. They attribute this overestimate of the oxygen abundance to high N/O and to low ionization. 

\subsubsection{The Ionization Parameter}

\begin{figure}
\includegraphics[width=\columnwidth]{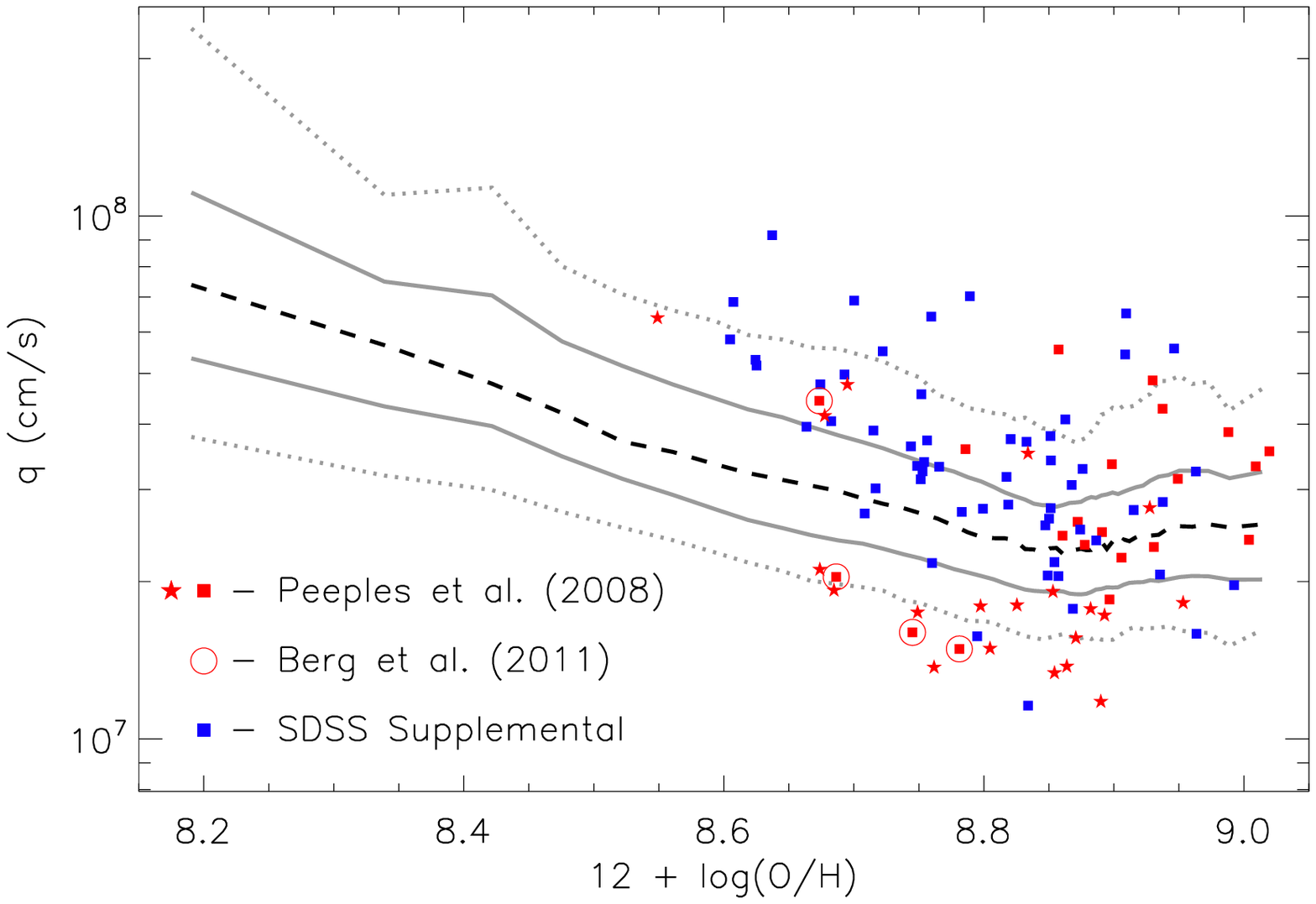}
\caption{The ionization parameter plotted against stellar mass. The \citet{Peeples2008} sample for which [OII]$\lambda$3727, 3729 is observed is plotted by the red squares. When [OII] is not observed, we use [SII] as a proxy. These galaxies are plotted by red stars. The 56 galaxies comprising our supplemental sample are plotted by the blue squares. The $O32$ value for SDSS sample taken from \citet{Zahid2011} is plotted by the dashed line. The data is sorted into 50 bins of stellar mass and the median $O32$ value is taken in each bin. The 68 and 95\% contours of the data are plotted by the solid and dotted gray curves.}
\label{fig:o32}
\end{figure}

We gain a handle on the ionization level of the gas by examining the ionization parameter (see Section 3.3 for derivation). In Figure \ref{fig:o32} we have plotted the ionization parameter as a function of stellar mass. The red points plot the sample from \citet{Peeples2008} with the four galaxies investigated by \citet{Berg2011} circled. For half of the \citet{Peeples2008} sample, the [OII] doublet is not observed. For these data we have used the [SII]$\lambda$6717, 6731 line as a proxy. The [SII] flux is strongly correlated to the [OII] flux owing to their similar ionization potential and primary origin. We derive a linear relation between the [SII] and [OII] dereddened line flux from $\sim$20,000 galaxies in the \citet{Zahid2011} sample and use this to infer the [OII] line flux when only [SII] is observed. The rms in the relation is 0.16 dex. These objects are shown by the red stars in Figure \ref{fig:o32}. The dashed black curve is the ionization parameter for the SDSS sample binned by stellar mass and the gray solid and dotted curves are the 68 and 95\% contours, respectively.

We are not able to ascertain the distribution of ionization parameter for our DEEP2 sample due to inability to measure the same line at two ionization stages (e.g. [OII] and [OIII]). It may be that our sample of DEEP2 galaxies have low ionization, though this is not a general feature of low mass, metal-rich galaxies. Future observations may shed light on this issue.

Consistent with the findings of \citet{Berg2011}, we find that three of the four galaxies that they investigated have ionization parameters that are lower than the larger SDSS sample. However, many of the galaxies in the supplemental SDSS sample (blue points) and the sample of \citet[][red points]{Peeples2008} appear to have ionization parameters consistent with the larger SDSS sample from \citet{Zahid2011}. This suggests that though low ionization may be an issue for some of the low mass, metal-rich galaxies, it is not generally the case. 

\subsubsection{Empirical Correction for Nitrogen Enhancement}

For our metal-rich sample we use the [NII]$\lambda6584$ line to measure the oxygen abundance. Nitrogen is known to have a primary component, formed mostly in massive stars, and a secondary component, formed in low and intermediate mass stars. Oxygen on the other hand is a primary element. Oxygen abundances inferred from $N2$ line ratio are known to depend on nitrogen to oxygen ratio (N/O) \citep{Storchi-Bergmann1994, Denicolo2002, Perez-Montero2009b}. The secondary production of nitrogen is dependent on the amount of oxygen already present in the star through the CNO cycle, producing a larger N/O at higher metallicity. Similarly, this dependence leads to a low dispersion in the N/O ratio at low metallicities \citep{Edmunds1978, Alloin1979} with the dispersion increasing at higher metallicities \citep{Perez-Montero2009b}.

\begin{figure}
\includegraphics[width=\columnwidth]{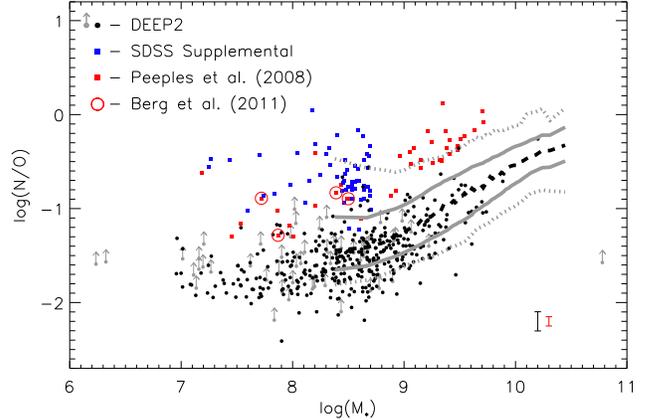}
\caption{N/O plotted as a function of stellar mass. The black points are 373 galaxies from the DEEP2 sample. For 59 galaxies in the DEEP2 sample, we observe [SII] with S/N $<$ 3. For these galaxies we have adopted a 3$\sigma$ limit for the [SII] EW. These data give a lower limit for N/O and are plotted by the gray arrows. The red and blue squares are the \citet{Peeples2008} and our supplemental sample, respectively. The four galaxies circled are the ones reexamined by \citet{Berg2011}. The median error for the nitrogen abundance of 0.1 and 0.05 dex for the DEEP2 and SDSS sample are shown by the black and red error bar in the bottom right corner, respectively. }
\label{fig:no}
\end{figure}

We determine N/O using Equation \ref{eq:no}. For the DEEP2 sample we do not have flux calibrated data, so we substitute in the equivalent widths of [NII] and [SII] lines when determining $N2S2$. \citet{Kobulnicky2003b} and \citet{Zahid2011} have demonstrated that substituting equivalent widths for line fluxes when measuring line ratios does not introduce significant errors. Figure \ref{fig:no} shows the N/O plotted as a function of stellar mass. The nitrogen abundances of the \citet{Peeples2008} and supplemental SDSS samples are elevated relative to the DEEP2 sample. This is consistent with the generally higher metallicities of the  \citet{Peeples2008} and supplemental SDSS samples as compared to the DEEP2 sample (see Figure \ref{fig:deep}) and due to metallicity dependency of secondary production, an independent measure of these galaxies having high metallicities. 

\begin{figure}
\includegraphics[width=\columnwidth]{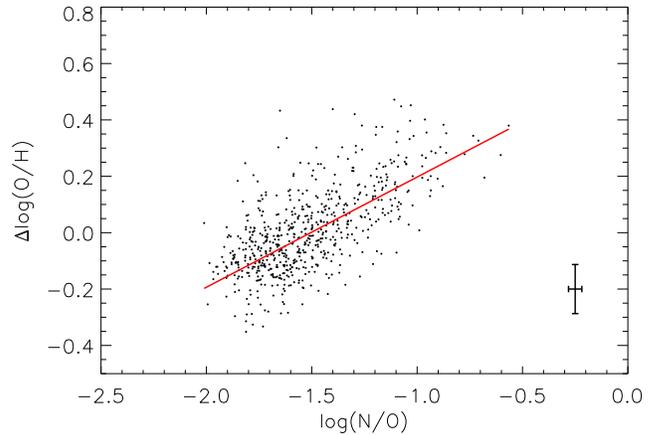}
\caption{The difference in metallicity between the strong line method and the direct method plotted against N/O. The sample is 627 galaxies from the SDSS DR7 with emission lines of interest detected with a S/N $>$ 5. The red line is a linear fit to the data taking into account errors in both coordinates. The median error in the metallicity difference and N/O is 0.09 and 0.03 dex, respectively.}
\label{fig:sdss_nocorr}
\end{figure}

We derive an empirical correction for enhanced nitrogen abundance by comparing the difference between direct method metallicities with those measured using the strong line method. We select all star-forming galaxies in the SDSS sample that have S/N $>$ 5 detections of [OII$]\lambda3727, 3729$,  [OIII]$\lambda4363, 4959, 5007$, H$\beta$ and [SII]$\lambda6717, 6731$. There are 627 galaxies that meet this criteria. We measure the oxygen abundance using the strong line calibration of \citet{Yin2007}, the direct method as parameterized by \cite{Izotov2006} and N/O using the calibration of \citet{Amorin2010}. In Figure \ref{fig:sdss_nocorr} we plot the difference in metallicity between the strong line method and the direct method as a function of N/O. The strong line method increasingly overestimates the metallicity with respect to the direct method as a function of N/O for this sample.

We have determined the errors in $\Delta$log(O/H) by adding in quadrature the errors from the strong line and direct method metallicities. The median error in  $\Delta$log(O/H) and N/O is 0.09 and 0.03 dex, respectively. The direct method metallicities dominate the errors in $\Delta$log(O/H) owing to the weakness of the [OIII]$\lambda4363$ line. We fit a linear relationship to the data using the MPFITEXY routine which takes into account errors in both coordinates \citep{Williams2010}. The MPFITEXY routine depends on the MPFIT package \citep{Markwardt2009}. The linear fit is given by
\begin{equation}
\Delta \mathrm{log(O/H)} = (0.64 \pm 0.03) + (0.42 \pm 0.02) \times \mathrm{log(N/O)}
\label{eq:nocorr}
\end{equation}
and is plotted by the red line in Figure \ref{fig:sdss_nocorr}. The rms of the fit to the data is 0.11 dex, with 0.07 dex attributable to the observational uncertainties.

\subsubsection{Correction Overestimates}

In making the correction for N/O, we have assumed that the direct method provides the most reliable metallicity estimate and therefore have corrected our strong line metallicities. We offer some words of caution in making this type of correction. The correction has been derived as a function of N/O. The galaxies used in deriving the empirical correction mostly lie in the region between $-2.0 < $ log(N/O)$\, < -1.0$ (see Figure \ref{fig:nocorr}). We have extrapolated the correction for galaxies with log(N/O) $> -1.0$, where most of the galaxies from the \citet{Peeples2008} and our supplemental SDSS sample lie.

In applying this correction, we have assumed that the direct method provides the most reliable metallicity measure. However, temperature fluctuations and gradients in HII regions may lead to underestimates of the metallicity with the direct method \citep{Stasinska2002, Stasinska2005, Bresolin2006}. The [OIII]$\lambda4363$ line strength increases with temperature. In the presence of fluctuations or gradients the inferred temperature may be biased towards higher temperatures in which case it would not be representative of the HII regions, especially in global spectra \citep{Kobulnicky1999}. 

\section{Scatter in MZ Relation at Low Masses}

\begin{figure*}
\includegraphics[width=2\columnwidth]{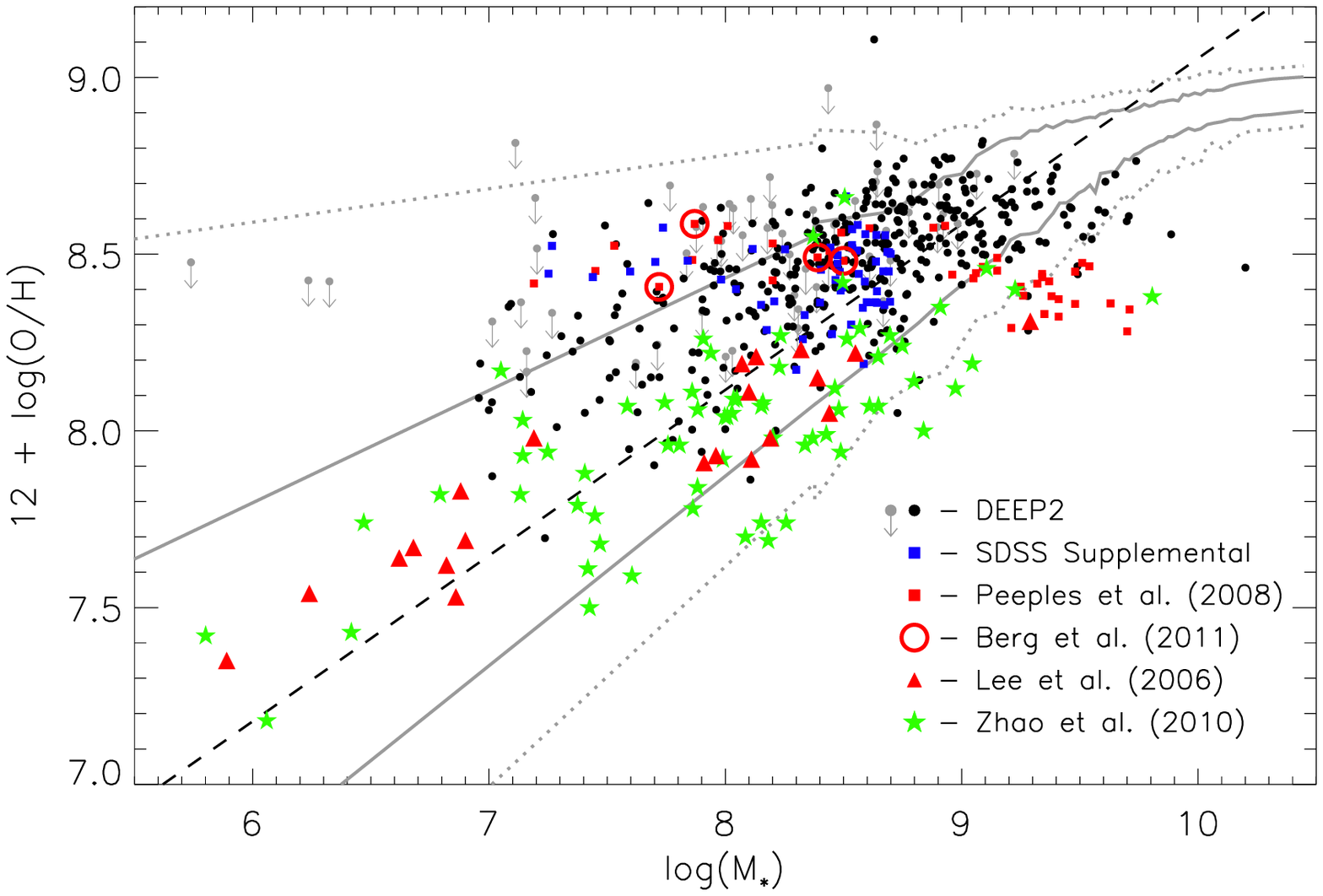}
\caption{The metallicity plotted against stellar mass. For the DEEP2  (black dots and gray arrows), \citet[][red squares]{Peeples2008} and supplemental (blue squares) sample we have determined metallicities using the $N2$ diagnostic. We apply an empirical correction to these data for enhanced nitrogen enrichment given in Equation \ref{eq:nocorr}. We also plot the samples of \citet[][red triangles]{Lee2006} and \citet[][green stars]{Zhao2010} for which metallicities have been determined using the direct method.}
\label{fig:nocorr}
\end{figure*}

In Figure \ref{fig:nocorr} we apply our derived empirical correction for enhanced nitrogen enrichment to the data. Even with the correction for enhanced N/O, low mass, metal-rich galaxies remain. The empirical correction for N/O brings the metallicities of the four galaxies investigated by \citet{Berg2011} down to 12 + log(O/H)$\sim$8.4-8.5 (circled galaxies in Figure \ref{fig:nocorr}). The low ionization in three of the four galaxies would further reduce the metallicity estimate. \citet{Berg2011} conclude that the metallicities of their four galaxies probably lie in the range of 7.9 $<$ 12 + log(OH) $<$ 8.4. Our analysis supports this conclusion. However, the same analysis applied to the rest of the galaxies in this study also supports the conclusion that \emph{not all of the low mass, metal-rich galaxies have overestimated metallicities}.

For galaxies above 10$^{9}\, \mathrm{M}_{\odot}$, Figure \ref{fig:nocorr} suggests that we may be overcorrecting for N/O. Because of this potential overcorrection and the fact that the sample is incomplete, the scatter at low stellar masses could only be \emph{larger} than what we observe. We consider Figure \ref{fig:nocorr} to display a lower limit to the intrinsic scatter in metallicities as a function of stellar mass. The observed scatter down to 10$^{7}$ M$_\odot$ is only slightly lower than what is inferred from the SDSS MZ relation.

\section{Physical Properties}

We explore some of the physical properties of the low mass galaxies and compare them with typical galaxies in the SDSS sample to understand their physical nature.

\subsection{Galaxy Colors, SFRs and Equivalent Widths}

\begin{figure}
\includegraphics[width=\columnwidth]{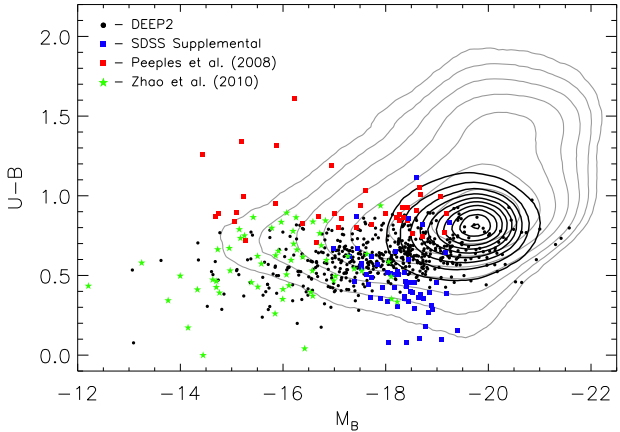}
\caption{The color-magnitude diagram of our sample of galaxies. The DEEP2 and \citet{Zhao2010} sample are plotted in black dots and green stars, respectively. The \citet{Peeples2008} and supplemental low mass, metal-rich sample from SDSS are plotted by the red and blue square, respectively. The black contours are for the main SDSS sample of star-forming galaxies taken from \citet{Zahid2011}. The gray contours for $\sim200,000$ galaxies in the SDSS.}
\label{fig:cmd}
\end{figure}

The color-magnitude diagram is shown in Figure \ref{fig:cmd}. The low mass galaxies have lower luminosities than the main sample of SDSS galaxies (black contours) and tend to be slightly bluer in color. The \citet{Peeples2008} galaxies (red squares) appear to be the reddest galaxies of all the samples. The metal-rich galaxies from the DEEP2 survey and the supplemental sample from SDSS appear to have colors that are consistent with the galaxies from the sample of \citet{Zhao2010}, which are blue compact dwarf galaxies selected on the basis of their color. The gray contours are for $\sim200,000$ galaxies from SDSS selected to be non-AGN using the BPT diagram and have $z<0.1$. The red sequence is shown by the locus of galaxies in the top right of the figure. Some of the galaxies of \citet{Peeples2008} have colors consistent with the red sequence, supporting the conclusion that these are transitionary objects.

\begin{figure}
\includegraphics[width=\columnwidth]{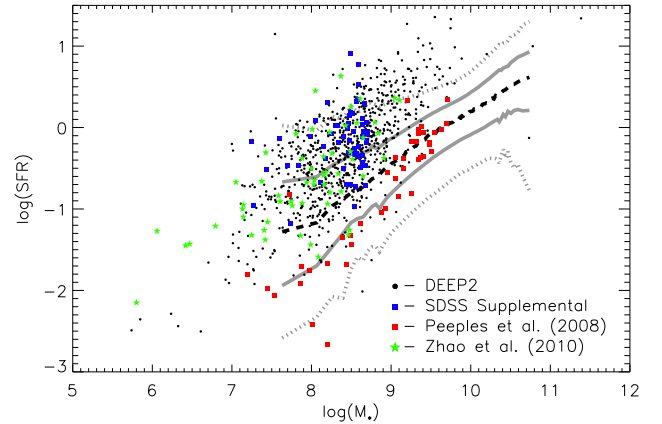}
\caption{The SFR plotted as a function of stellar mass. The DEEP2 and \citet{Zhao2010} sample are plotted in black dots and green stars, respectively. The \citet{Peeples2008} and supplemental low mass, metal-rich sample from SDSS are plotted by the red and blue square, respectively. The black dashed curve is the median SFR in 100 bins of stellar mass for $\sim140,000$ star-forming galaxies in the SDSS. The 68 and 95\% contours are shown by the solid an dotted gray curves, respectively}
\label{fig:sfr}
\end{figure}

In Figure \ref{fig:sfr} we examine the SFRs as a function of stellar mass. High mass galaxies tend to have higher SFRs. The SFRs of \citet{Peeples2008} sample generally are lower than the other samples, consistent with their redder colors. The distribution of SFRs for the DEEP2, \citet{Zhao2010} and supplemental SDSS samples appear to be similar. The black dashed curve is the SFRs for $\sim140,000$ star-forming galaxies in the SDSS sorted into 100 stellar mass bins. These data are selected to have a S/N$>$5 in H$\alpha$ and H$\beta$, $z<0.1$ and are required to be classified as star-forming in the BPT diagram. In general, the data used in this study appeare to be biased towards higher SFRs relative to the local galaxies from SDSS. For the DEEP2 data, this is partly a result of redshift evolution in the SFRs as these data have $z \lesssim 0.4$. This plot illustrates some of the bias in the samples used in this study. A more complete sample is required to probe galaxies with lower SFRs.

\begin{figure}
\includegraphics[width=\columnwidth]{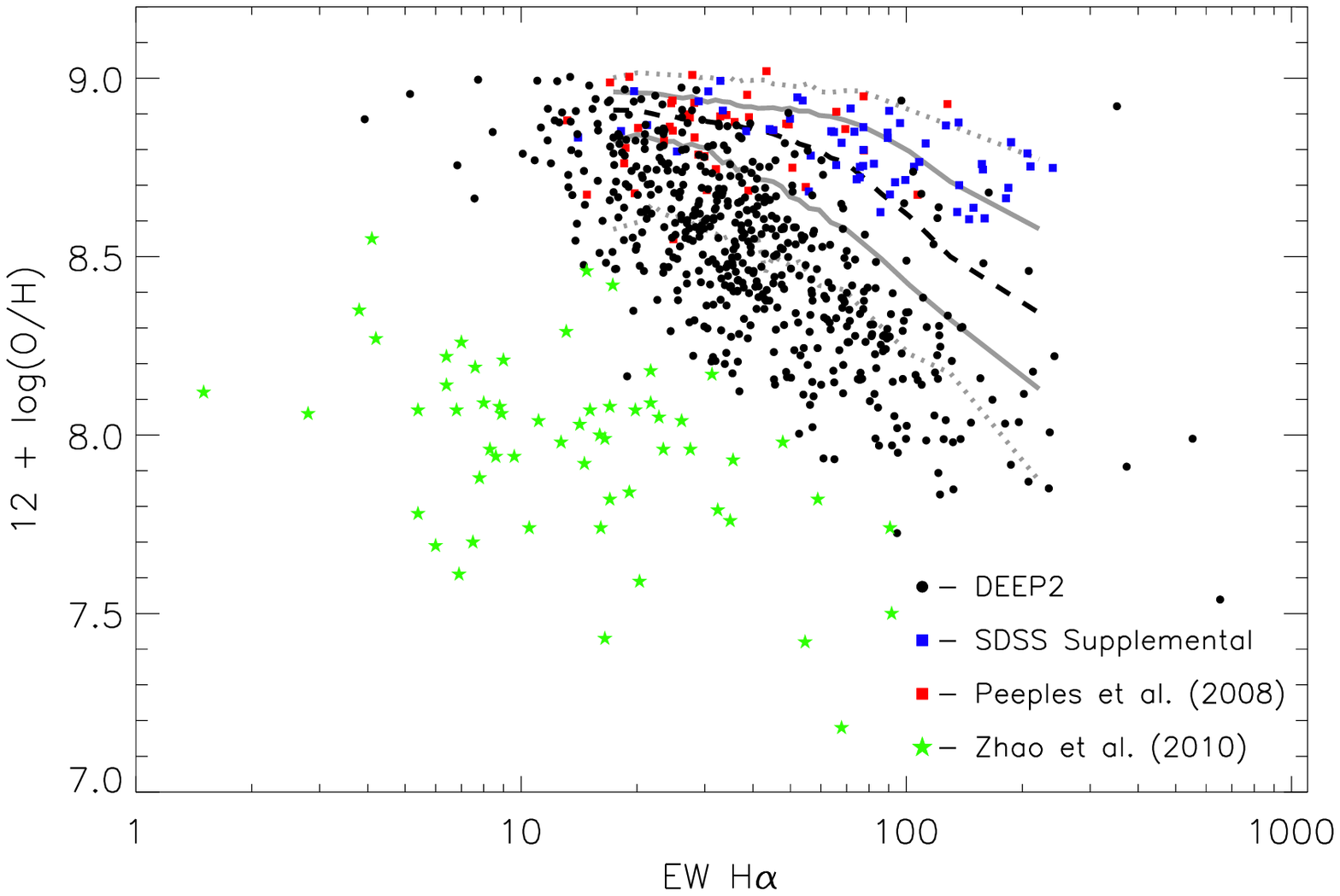}
\caption{The metallicity as a function of the equivalent width of H$\alpha$ (EW H$\alpha$). The DEEP2 and \citet{Zhao2010} sample are plotted in black dots and green stars, respectively. The \citet{Peeples2008} and the supplemental low mass, metal-rich sample from SDSS are plotted by the red and blue square, respectively. The median metallicity in 50 bins of EW H$\alpha$ for the SDSS sample taken from \citet{Zahid2011} is plotted by the dashed black curve. The 68 and 95\% contours are shown by the solid an dotted gray curves. The equivalent width is observed to strongly decrease with increasing metallicity. }
\label{fig:ewha}
\end{figure}

In Figure \ref{fig:ewha} we plot the metallicities of our galaxies against the equivalent width of H$\alpha$ (EW H$\alpha$). Here we have not applied our correction for elevated N/O as this may be an overestimate and does not affect the interpretation. The EW H$\alpha$ is a measure of the H$\alpha$ line flux divided by the underlying continuum. The flux of H$\alpha$ is a proxy of star formation and the underlying continuum is dominated by the emission of the older, low and intermediate mass stars. This population probed by the underlying continuum of H$\alpha$ dominates the stellar mass of a galaxy. The EW H$\alpha$ represents a measure of the amount of star formation normalized to stellar mass and therefore changes with time. We choose to plot EW H$\alpha$ instead of specific star formation rate because it is a directly observable quantity and is not subject to the uncertainties in stellar mass and star formation rate. There appears to be a trend in each of the samples separately such that galaxies with lower EW H$\alpha$ tend to have higher metallicities. 

\subsection{Physical Nature of Low Mass Galaxies}

The blue colors observed the low mass galaxies are typical of dwarf galaxies. \citet{Schombert1995} find that their sample of dwarf spiral galaxies have colors that are bluer than the normal early-type spirals despite their low star formation rates \citep[see also][]{Hidalgo-Gamez2004}. These colors are also consistent with those observed in quiescent gas-rich dwarf galaxies \citep{VanZee1997}. \citet{Peeples2008} sample of galaxies are, in general, the reddest galaxies. This is consistent with the lower SFRs observed in those galaxies.

Blue colors are a general property observed in star-forming galaxies. Blue colors imply a population of young, high mass stars which tend to dominate the luminosity output of a galaxy. The fact that blue colors are observed in these dwarf galaxies even though many exhibit low levels of star formation as inferred from the H$\alpha$ line luminosity suggests that even low levels of localized star formation in low stellar mass systems can dominate the color. \citet{Peeples2008} observe that many of their galaxies are dominated by blue cores. In the supplemental SDSS sample, a few galaxies also have blue cores, though in the majority of galaxies star formation appears to be more widespread.

One of the most interesting features of our sample of galaxies is the relation between metallicity and EW H$\alpha$ as seen in Figure \ref{fig:ewha}. A similar trend is observed in the DEEP2, SDSS and \citet{Zhao2010} sample such that the highest metallicity objects in each sample have the lowest EWs. However, though the three samples have a similar range in EWs, they are somewhat segregated in metallicity such that they cluster into what appears to be distinct populations. In Figure \ref{fig:ewha}, there appears to be a continuum between the DEEP2 and SDSS data, but a gap in metallicity between the \citet{Zhao2010} sample. We hypothesize that a complete census of low mass galaxies cover the full range of observed metallicities and EW H$\alpha$.
 
The low and high metallicity sample of galaxies investigated in this study have similar masses, colors, SFRs and EW H$\alpha$. This implies that these galaxies likely have similar stellar populations. Under this assumption, the most straight-forward interpretation of Figure \ref{fig:ewha} would be that the differing metallicities in the samples is due to the gas content of the galaxies, such that the lower metallicity galaxies at a given EW H$\alpha$ are more gas rich. This is consistent with the interpretation of \citet{Dellenbusch2007} and \citet{Peeples2008}.

\section{Discussion}

Understanding and quantifying the scatter in the MZ relation is crucial to uncovering the origin of the MZ relation. Observations of low mass, metal-rich galaxies help shed light on this issue but also present us with a recently discovered population of galaxies requiring further investigation and characterization.

The MZ relation at low stellar masses has only been studied by a few authors \citep{Tamura2001, Lee2006, Vaduvescu2007, Petropoulou2011}. These studies have been hampered by a small number of observations owing to the faintness of low stellar mass galaxies and selection biases. \citet{Lee2006} conclude that MZ relation extends to lower metallicities with no increase in the scatter. We have shown that the scatter in the MZ relation increases substantially ($\sim$0.4 dex) over the 2 decades of stellar mass observed in the SDSS MZ relation (Figure \ref{fig:scatter}).

By examining data from the DEEP2 and SDSS surveys, we are able to identify low mass, metal-rich galaxies that appear to be consistent with the increasing scatter implied by the SDSS MZ relation (Figure \ref{fig:mzr}). The observations of low mass, metal-rich galaxies challenge the notion that the MZ relation extends to low stellar masses with no increase in the scatter as found by \citet{Lee2006}. Furthermore, given their low stellar masses and shallow potential wells implied by the velocity dispersion (see Figure \ref{fig:vdisp}), it is difficult to understand these galaxies in the context of the canonical explanation for the MZ relation which argues that shallow potential wells in lower mass galaxies prevent them from enriching due to loss of metals through galactic scale outflows \citep[e.g.][]{Tremonti2004}. 

The MZ relation is observationally well established in both the local \citep[e.g.][]{Tremonti2004} and high redshift universe \citep[e.g.][]{Erb2006b, Mannucci2009}. The MZ relation shows that the median (or mean) metallicity of star-forming galaxies increases as a function of stellar mass. Our results suggest that in the local universe this is largely driven by a more rapid decline in the metallicity of the most metal-poor galaxies relative to metal-rich galaxies at lower stellar masses (see Figure \ref{fig:deep} or \ref{fig:nocorr}). This observed trend puts stronger constraints on theoretical chemical evolution models attempting to reproduce the MZ relation.

In Figure \ref{fig:scatter}, we have compared the scatter in the MZ relation from several diagnostics with the theoretical scatter taken from the smoothed particle hydrodynamical simulations of \citet{Dave2011b, Dave2011a}. \citet{Dave2011b} investigate the MZ relation using several prescriptions for winds. Their simulations include a no wind model, two constant wind models (with velocities of 340 and 680 km s$^{-1}$) and a momentum driven wind such that the wind speed scales with the velocity dispersion of the galaxy. The constant wind models have considerably higher scatter at low stellar masses than the observations and the no wind model generally shows less scatter than the observations. The momentum driven wind model best characterizes the observational data and the scatter predicted by this model is shown in Figure \ref{fig:scatter}. The overall normalization is a free parameter in the models but in addition to the scatter, the shape and slope of the local MZ relation \citep[from][]{Tremonti2004} are also well fit by this model \citep[see their Figure 1]{Dave2011b}. The predicted scatter in the models deviates most strongly at lower stellar masses. Therefore, the scatter in the MZ relation at lower masses will provide important tests for theoretical models of chemical evolution. However, current simulations \citep[e.g.][]{Dave2011b, Dave2011a} do not resolve galaxies with stellar masses below 10$^9 \, \mathrm{M}_\odot$ and we await higher resolution simulations to make this comparison.

From their simulations, \citet{Dave2011b} argue that galaxies evolve around an equilibrium MZ relation. Large scale gas flows perturb galaxies from equilibrium and the scatter in the MZ relation is a quantitative measure of the time it takes to return to equilibrium which, in their simulations, is reestablished through infall of gas. They argue if the dilution time of the gas (gas mass divided by inflow rate) is small compared to the dynamical time of the galaxy (at the virial radius), then the scatter will be small. Within this framework, our observations suggest that the timescales for galaxies to equilibrate once they are perturbed are longer at lower stellar masses (i.e. the dilution time is longer than the dynamical time at lower stellar masses).

In the local universe, the SFRs of star-forming galaxies are (anti-)correlated to the metallicity in the sense that the higher metallicity objects at a given stellar mass have lower SFRs \citep{Ellison2008, Mannucci2010, Dave2011b}. For our low mass samples, this trend is only weakly present in the data (see Figure \ref{fig:sfr}). Specifically, the scatter in the MZ relation at low stellar masses can be \emph{slightly} reduced if SFRs are accounted for in the manner presented in \citet{Mannucci2010}. However, the SDSS supplemental sample and some of the DEEP2 galaxies appear to have both high metallicities and SFRs, inconsistent with the observed trend at higher stellar masses. The full implication of this correlation is still not well understood, though it is probably related to the gas content which regulates both the gas-phase abundance and the SFR. The fact that this effect is relatively weak in lower mass galaxies suggests that the relative contribution of various physical mechanisms governing the metallicities of low mass galaxies may differ from higher mass galaxies.

Our analysis supports the conclusion that the sample of low mass, metal-rich galaxies appear to be ``transitional" dwarf galaxies nearing the end of their star formation \citep{Grebel2003, Peeples2008}. This interpretation suggests that metal-rich galaxies have low gas fractions relative to other dwarf galaxies of similar stellar mass and the low gas fractions are largely responsible for the high gas-phase oxygen abundance observed. \citet{Dellenbusch2007} come to similar conclusions for their smaller sample of low luminosity, metal-rich galaxies. In this evolutionary scenario, these galaxies would represent the link between the gas-rich dwarf irregulars and the gas-deficient dwarf spheroidals and ellipticals. The bluer galaxies in the supplemental SDSS sample would be at an earlier evolutionary stage in this transition, where star formation is more widespread than the \citet{Peeples2008} sample but gas content is lower than their low metallicity counterparts. Accurate determinations of the gas content of low mass, metal-rich galaxies are important to test and establish this scenario.

The environment of these low mass galaxies may play a crucial role in determining their chemical properties. Both observations and simulations have suggested that galaxies in dense environments tend to have metallicities over and above those expected from the correlation of stellar mass and the environment \citep{Cooper2008, Dave2011b}. Moreover, the stripping away of gas from low mass galaxies in high density environments, either through ram pressure stripping or interactions, has been cited as the possible explanation for high metallicities \citep{Boselli2008, Petropoulou2011}. However, \citet{Peeples2008} conclude that their galaxies are fairly isolated and the DEEP2 and supplemental SDSS sample of galaxies, at least in projection, do not appear to be in high density environments though further investigation into the environments of low mass, metal-rich galaxies is necessary.

In order for ram pressure stripping to occur effectively it is estimated that $P_{ram} \sim \rho_{\mathrm{IGM}} v^2 > \sigma^2 \rho_{\mathrm{gas}}/3$ \citep{Gunn1972}. Here, $P_{\mathrm{ram}}$ is the ram pressure, $\rho_{\mathrm{IGM}}$ is the gas density of the intergalactic medium (IGM), $v$ is the velocity of the galaxy through the IGM, $\sigma$ is the velocity dispersion and $\rho_{\mathrm{gas}}$ is the gas density of the galaxy. Typical IGM densities ($n_\mathrm{H} < 10^{-5}$ cm$^{-3}$) are likely not sufficient to remove significant amounts of gas. However, as \citet{Grebel2003} argue, a combination of tidal effects along with ram pressure stripping as dwarf galaxies pass near giant galaxies or dense regions of the IGM (if the medium is inhomogeneous) is the most plausible mechanism for gas removal. In this scenario, even if dwarf galaxies are not currently found in high density environments, previous encounters could remove gas from the ISM and any subsequent star formation would lead to elevated levels of enrichment due to the low gas fractions.

The blue cores and low SFRs of many of the metal-rich galaxies suggests that star formation is not widespread throughout the galaxy. Localized star formation opens up the possibility that the high metallicities observed in some of these galaxies may be a result of the nebular emission dominated by individual HII regions. In this case, the inferred metallicity may not be indicative of the global metallicity of the galaxy. \citet{Gu2006} were the first to observe a metal-rich dwarf galaxy. From detailed photometric and spectral analysis of IC 225, they find that there are two photometric cores in this galaxy that are spatially separated; the off-nuclear core being bluer than the blue nucleus. From the redshifts of the line emission and absorption, they conclude that the line emission originates from the off-nuclear core. Follow up kinematic study of the nuclear region of IC 225 using integral field spectroscopy have confirmed this result \citep{Miller2008}. 

In most star-forming galaxies, the gas-phase oxygen abundance is determined from emission lines integrated over many star-forming regions and hence considered to reflect the global metallicity. If, in these low mass systems, the emission lines are dominated by a single region, the metallicity determined from these lines may not reflect the global gas-phase abundance. Spatially resolved spectroscopic studies of the metal-rich dwarf galaxies will help to establish whether the high metallicities observed are representative of global metallicities or are localized to individual HII regions. If the latter is true, determining at what stellar masses this effect becomes pronounced and the systematic effects this has on metallicity studies of galaxies is crucial.

\section{Summary}

In this contribution, we have investigated the metallicities of low mass galaxies. Our data comes from culling the literature and from identifying low mass galaxies in the SDSS and DEEP2 survey. We summarize our main findings as follows:

\begin{enumerate}

\item{We examine the scatter in the local stellar mass-metallicity relation determined from $\sim$20,000 galaxies in the SDSS. We find that the scatter increases with decreasing stellar mass. We observe this trend using several diagnostic methods of metallicity determination mitigating the systematic effects and uncertainties associated with any particular diagnostic. The theoretical scatter taken from smoothed particle hydrodynamical simulations with momentum driven winds is also examined and found to be consistent with the observed scatter (see Figure \ref{fig:scatter}).}

\item{The low stellar mass MZ relation derived from galaxies with metallicities determined using the direct method shows a smaller range of metallicities than the SDSS MZ relation. We attribute this lower observed scatter to selection bias in galaxies with metallicities determined by the direct method. Using the $N2$ strong-line diagnostic method, we provide observations of low mass, metal-rich galaxies. We examine systematic uncertainties of enhanced nitrogen enrichment and variations in the ionization parameter to help establish the metal-richness of these low mass galaxies. The results  challenge the notion that the scatter in the low stellar mass-metallicity relation is small and constant with stellar mass. Due to incompleteness of our sample and the possibility that our empirical correction for elevated N/O is overcorrecting the data, we give a lower limit to the scatter at low stellar masses (see Figure \ref{fig:nocorr}).}


\item{Low mass, metal-rich objects have been identified in the literature as the transitional objects between gas rich dwarf irregulars and gas poor dwarf spheroidals and ellipticals \citep{Dellenbusch2007, Peeples2008}. We examine the physical properties of these low mass galaxies and find that they are generally consistent with this scenario. }

\end{enumerate}

Further observations are required to understand these recently discovered population of galaxies. In particular, accurate gas mass determinations will help to establish these low mass, metal-rich galaxies as gas poor. Spatially resolved spectroscopy of individual galaxies will help to determine if the metallicities inferred from the emission lines are indicative of the global metallicities. The effects that nitrogen enhancement and low ionization parameter have on metallicities are not well understood. A firm observational basis is required for deriving empirical corrections. Finally, detailed investigations into the environments of these galaxies may shed light on physical mechanisms leading to their ostensibly high metallicities and their roles within the sequence of galaxy evolution.

\acknowledgments

HJZ and LJK gratefully acknowledge support by NSF EARLY CAREER AWARD AST07-48559. F.B. gratefully acknowledges the support from the National Science Foundation grants AST-0707911 and AST-1008798. We thank Kevin Bundy for generously sharing his K-band photometry and Yinghe Zhao for providing their data. We also like to thank Ben Weiner, Dan Weisz, Danielle Berg for useful discussion and Danielle Berg for providing their line flux measurements that were not published. We are grateful to Stephane Arnout and Olivier Ilbert for making their photo-z code available for use in estimating galaxy stellar mass. We acknowledge the cultural significance Mauna Kea has for the Hawaiian community and with all due respect say mahalo for its use in this work.

We thank the DEEP2 team for making their data publicly available. The analysis pipeline used to reduce the DEIMOS data was developed at UC Berkeley with support from NSF grant AST-0071048. 

Funding for the SDSS and SDSS-II has been provided by the Alfred P. Sloan Foundation, the Participating Institutions, the National Science Foundation, the U.S. Department of Energy, the National Aeronautics and Space Administration, the Japanese Monbukagakusho, the Max Planck Society, and the Higher Education Funding Council for England. The SDSS Web Site is http://www.sdss.org/.

The SDSS is managed by the Astrophysical Research Consortium for the Participating Institutions. The Participating Institutions are the American Museum of Natural History, Astrophysical Institute Potsdam, University of Basel, University of Cambridge, Case Western Reserve University, University of Chicago, Drexel University, Fermilab, the Institute for Advanced Study, the Japan Participation Group, Johns Hopkins University, the Joint Institute for Nuclear Astrophysics, the Kavli Institute for Particle Astrophysics and Cosmology, the Korean Scientist Group, the Chinese Academy of Sciences (LAMOST), Los Alamos National Laboratory, the Max-Planck-Institute for Astronomy (MPIA), the Max-Planck-Institute for Astrophysics (MPA), New Mexico State University, Ohio State University, University of Pittsburgh, University of Portsmouth, Princeton University, the United States Naval Observatory, and the University of Washington.

\section*{Appendix}
\setcounter{secnumdepth}{-1}
\subsubsection{A1. Comparison of N2 Diagnostics}

\begin{figure}
\includegraphics[width=\columnwidth]{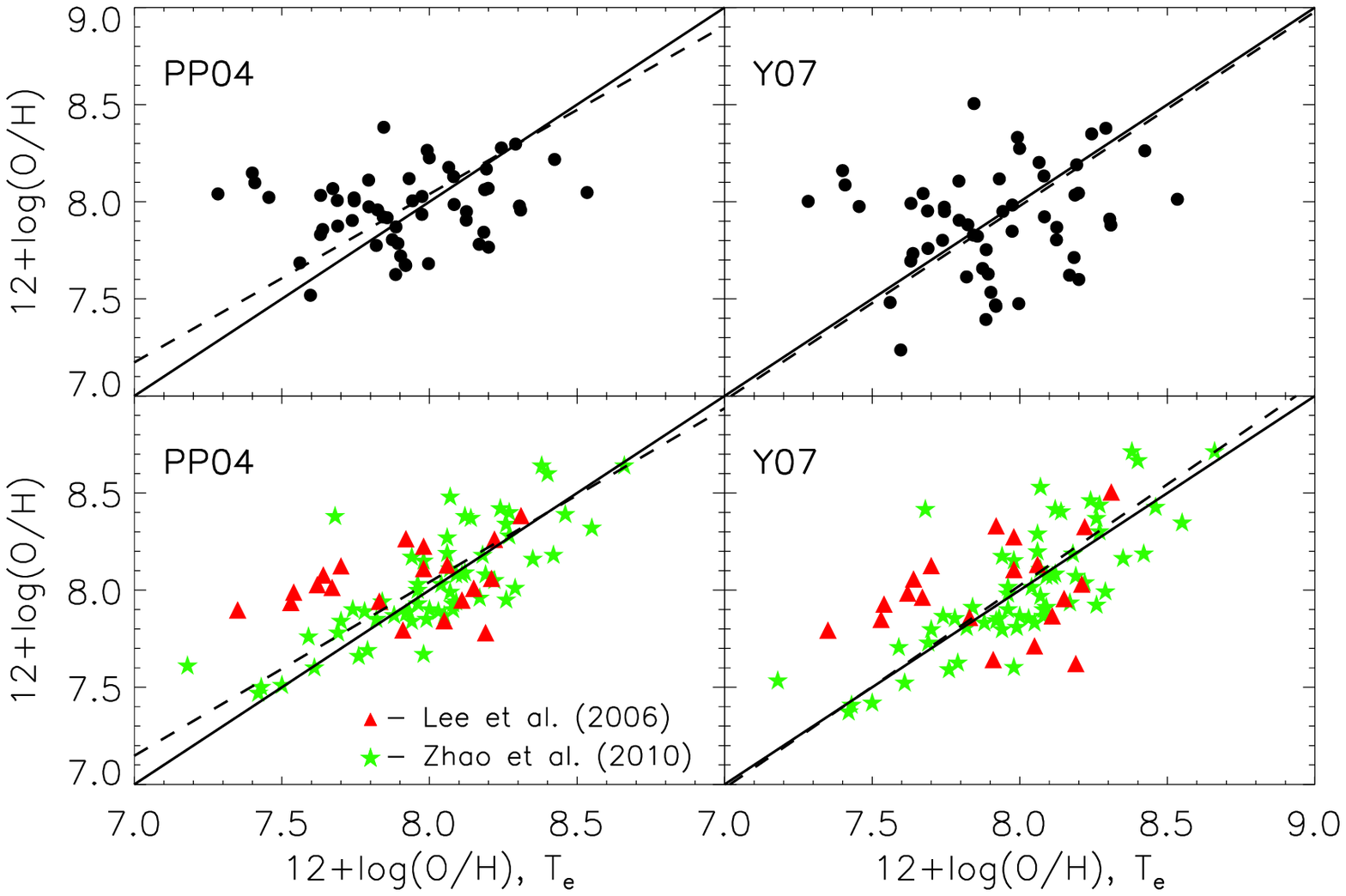}
\caption{A comparison of the $N2$ metallicity diagnostic with the direct method determination using the [OIII]$\lambda4363$ temperature sensitive line \citep{Izotov2006}. We have determined the metallicities using the strong line diagnostics calibrated by \citet[][left]{Pettini2004} and \citet[][right]{Yin2007}. The top panels show metallicities derived for individual HII regions. The bottom panels show global metallicities. The red triangles are 20 dwarf irregular galaxies from \citet{Lee2006}. For these galaxies we sum individual spectra for a given galaxy to obtain an integrated measure. The green stars are 60 compact blue galaxies with metallicities taken from \citet{Zhao2010} and the $N2$ values are taken from their paper. The solid line in each plot is the one-to-one agreement and the dashed line is a linear bisector fit.}
\label{fig:hii}
\end{figure}

In Figure \ref{fig:hii}, we compare the diagnostics of \citet[left panels]{Pettini2004} and \citet[right panels]{Yin2007}. In the top panels we compare data compiled from the literature taken from the objects investigated by \citet{Lee2006}. The compiled data are emission line measurements for individual HII regions. Only HII regions where the [OIII]$\lambda4363$, [NII]$\lambda6584$ and H$\alpha$ lines are measured are included. In the bottom panels, the comparison between the methods is made for galaxies taken from the samples of \citet{Lee2006} and \citet{Zhao2010}. For the \citet{Lee2006} sample, the integrated flux is determined by summing the flux from individual HII regions. The T$_e$ metallicities were calculated using the iterative scheme presented by \citet{Izotov2006}.

In Figure \ref{fig:hii}, the solid line in each plot is the one-to-one agreement. The data are independent and have errors in both the x and y direction. We perform a linear bisector fit using the routine $\textit{robust\_linefit.pro}$ in IDL in order to assess how well the data derived from the diagnostics agree with the direct metallicity determination. The dashed lines are a linear bisector fit in each case. A scatter between the relation of the direct and strong-line method of 0.30, 0.35, 0.23 and 0.24 dex are observed in the top left, top right, bottom left and bottom right panels, respectively. The greater slope in the \citet{Yin2007} calibration as compared to the \citet{Pettini2004} calibration leads to a slightly larger scatter (compare Equations \ref{eq:met_pettini} and \ref{eq:met_yin}). Though the \citet{Pettini2004} metallicities match the direct determinations reasonably well, the \citet{Yin2007} metallicities provide a more consistent calibration at lower metallicities. 

\subsubsection{A2. Conversion of R23 to T${_e}$ Metallicities}

\begin{figure}
\begin{center}
\includegraphics[width=\columnwidth]{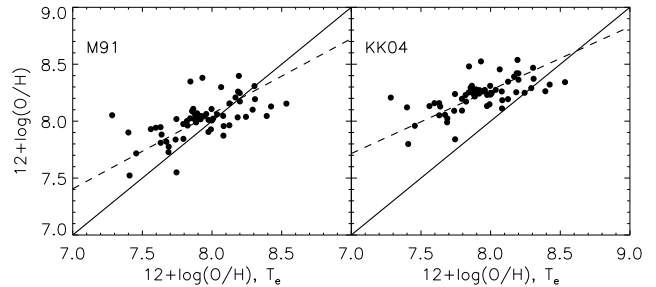}
\caption{The metallicities determined from $R23$ compared with the direct method. The left and right panel show the metallicity determined using the \citet{McGaugh1991} and \citet{Kobulnicky2004} metallicity diagnostics, respectively. The solid line is the one-to-one correspondence and the dashed line is a linear bisector fit to the data.}
\label{fig:r23}
\end{center}
\end{figure}

We give the conversions required to make metallicities from two commonly used calibrations of $R23$ consistent with the direct method at low metallicities. The left and right panel of Figure \ref{fig:r23} show the metallicities for the same HII regions as Figure \ref{fig:hii} determined from the calibrations of \citet{McGaugh1991} and \citet{Kobulnicky2004}, respectively. The relatively low metallicities of these HII regions place them on the lower metallicity branch of the $R23$ diagnostic. The slope of the linear bisector fit is 0.67 and 0.57 for the \citet{McGaugh1991} and \citet{Kobulnicky2004} calibrations respectively. 

At low metallicities $R23$ diagnostics present some advantages over $N2$. The reduced scatter in the $R23$ metallicity calibration as compared to the $N2$ calibration is due to observational uncertainty. At lower metallicities, the oxygen lines are significantly stronger than the [NII]$\lambda6584$ line, whose line strength diminishes with decreasing metallicity. This places some limitations on the usefulness of the $N2$ diagnostic for estimating metallicities in high redshift galaxies where we expect less enrichment and therefore lower metallicities. 

In order to make the metallicity found for the lower branch of the $R23$ diagnostic consistent with the direct method for our sample of low metallicity HII regions, we provide the following conversion:
\begin{equation}
\mathrm{M91}^\prime = 1.60 \times \mathrm{M91} - 4.87 \\
\end{equation}
\begin{equation}
\mathrm{KK04}^\prime = 1.90 \times \mathrm{KK04} - 7.70.
\end{equation}
Here, M91 and KK04 are the metallicities determined using the diagnostics of \citet{McGaugh1991} and \citet{Kobulnicky2004}, respectively. The prime quantities are those diagnostics linearly converted to be consistent with the direct method metallicities determined from the iterative scheme of \citet{Izotov2006}. This is also consistent with the $N2$ line ratio metallicity calibration determined by \citet{Yin2007}. We note that these conversions should only be used when the metallicities are known to be on the lower metallicity branch of $R23$.

\bibliographystyle{apj}

\bibliography{metallicity}
 \end{document}